# Mt. Wendelstein Imaging of the Post-Perihelion Dust Coma of 67P/Churyumov-Gerasimenko in 2015/2016

Hermann Boehnhardt (Max-Planck Institute for Solar System Research, Justus-von-Liebig-Weg 3, 37077 Göttingen, Germany; Email: boehnhardt@mps.mpg.de), Arno Riffeser, Matthias Kluge, Christoph Ries, Michael Schmidt, Ulrich Hopp (University Observatory, Ludwig-Maximilian-University Munich, Scheiner Str. 1, 81679 München, Germany)

**Correspondence:** Hermann Boehnhardt, MPI for Solar System Research, Justus-von-Liebig-Weg 3, D-37077 Göttingen, Germany; Email: boehnhardt@mps.mpg.de

**Abstract:** Comet 67P/Churyumov-Gerasimenko (67P) was imaged with the 2m telescope at Mt. Wendelstein Observatory in the Alps. Coma and tail monitoring was performed during 51 nights between 22 August 2015 and 9 May 2016. The images through r and i Sloan Digital Sky Survey (SDSS) filters show the dust distribution around the comet, while images in the SDSS g filter indicate also the presence of coma gas in early September 2015. The dust color of 67P implies intrinsic reddening of 9 %/100nm. After maximum shortly after perihelion passage the dust activity decreased with a heliocentric exponent of 4.1 to 4.2 from late September 2015 until May 2016. The opposition surge during early 2016 can be explained by a linear light scattering phase function ($\beta \sim 0.04$) or an asteroid-like HG-type phase function ($G \sim 0.15$). The radial brightness profile indicates a 'quasi-steady-state' dust coma from late September to the end of 2015. Dust fragmentation during about a month after perihelion may be responsible for radial coma profiles with slopes below unity, while dust accumulation due to very slow dust expansion velocity may result in steeper than unity profiles during 2016. Three fan-shape dust structures are characterized in the coma of 67P. A short dust ejection event on 22 -23 August 2015 has produced a dust arc-let and jet feature in the coma. In September 2015 the appearance of cometary dust tail is dominated by young dust produced around perihelion. The older dust dominates the tail appearance as of mid November 2015.

# 1. Introduction

In 2015 comet 67P/Churyumov-Gerasimenko (67P), a Jupiter family comet (JFC) and the main science target of ESA's Rosetta mission, had its 8th return to perihelion since its discovery on 9 September 1969 at the Alma Ata Observatory. Since 2003, when ESA selected 67P as the main target for Rosetta, the comet received much more attention by regular telescopic observations from Earth and in space in preparation and support of the mission (Schulz et al. 2004, Weiler et al. 2004, Lara et al. 2005+2011, Kelley et al. 2006+2008,+2009, Lamy et al. 2006+2008, Schleicher 2006, Ishiguro 2008, Tubiana et al. 2008+2011, Agarwal et al. 2010, Hadamcik et al. 2010, Tozzi et al. 2011, Lowry et al. 2012, Vincent et al. 2013, Snodgrass et al. 2013+2016). Furthermore, useful collections of ground-based observations were published by Kidger (2004) and Ferrin (2005). The observational results represented valuable input information for the Rosetta mission planning. Since for 67P they cover different aspects of cometary properties and a wider and denser sampled time window than for many other JFCs, they allow also studies of the time evolution of the comet as well as new interpretations and insights on the background of the results evolving from the Rosetta mission.

Here, we present observations and results from the 2015/2016 return of comet 67P. The imaging observations focus on the coma structure and dust production versus time and try to link the results to the knowledge on the comet from previous apparition as well as to recent conclusions from Rosetta data.

# 2. Observations and Data Reduction

The observations of 67P described here, were performed at the Mt. Wendelstein Observatory (12°00'45"E, 47°42'13"N, 1837 m above sea level) of the University of Munich from 22 August 2015 to 9 May 2016, i.e. covering the period from 9 to 270 days after perihelion passage of the comet. In this period the comet was imaged during 51 nights through Sloan Digital Sky Survey (SDSS; Fukugita et al. 1996) r' filter using the Wendelstein Wide Field Imager (WWFI) camera (Gössl et al. 2012) at the 2m Fraunhofer telescope (Hopp et al. 2014). During 4 nights the comet was also imaged through SDSS filters g' and i'. The WWFI camera has a f/7.8 optics and is equipped with a CCD mosaic from Spectral Instruments Tucson, mounting four 4k by 4k 15 μm 4-port read-out e2V CCDs that provides a pixel size of 0.2 arcsec and covers 27 x 27 arcmin on the sky. Information on the observing site and the technical equipment used can be found in Kosyra et al. (2014).

The Mt. Wendelstein imaging campaign of 67P covers the post-perihelion solar distance range of the comet from 1.248 to 2.966 AU with Earth distances between 1.491 (21 February 2016) and 2.386 AU (9 May 2016). During the campaign, the phase angle of the comet decreased first slowly from 33.9 deg at the beginning of the campaign to 29.3 deg on 29 December 2015, thereafter more rapidly to 4.0 deg on 14 March 2016 and increasing again after opposition to 17.8 deg at the end of the observing campaign. The pixel size of the camera resulted in projected extensions of 216.3 to 346.1 km at the comet. Tab. 1 provides the observing log of the 67P campaign at Mt. Wendelstein Observatory.

During the nightly observing window of the comet, image series of 67P were obtained with exposure times between a few 10sec (early phase of the campaign) up to 120sec (late phase) integration times, applying sidereal tracking and guiding to the telescope. The exposure times were chosen such to freeze the cometary motion to the typical seeing of the site. Multiple consecutive images were taken each night for improvement of the signal-to-noise ratio and to allow for the removal of background object as well as artificial features (e.g. bad pixels, cosmics, inter-chip regions) in the images through post-processing of the data. For the latter reason a fixed offset pattern (with 13 pointings) with a maximum amplitude of typically +/- 120 arcsec was applied from image to image during the nightly exposure series. The offset pattern took care of keeping the coma and the tail region close to the comet head well-placed inside the camera field of view.

**Table 1** Observing log of the comet 67P observations at the Mt. Wendelstein Observatory. All observations were performed with the WWFI camera at the Fraunhofer 2m-telescope. The table columns list the Modified Julian Date (MJD) and the civil observing date (UT), the observing time (UT), the Sun (r in AU) and Earth (Δ in AU) distance, the phase angle (deg) of the comet, the pixel scale (km) at the distance of the comet, the exposure time used (s), the number of exposures taken, the total integration time on comet (s), the WFFI filter used and remarks.

| MJD | Date (UT) | Time (UT) | r (AU) | Δ (AU) | Phase (deg) | Pix. Res. (km) | Exp. Time (s) | Exp. Number | Int. time (s) | WWFI Filter | Remarks |
|---|---|---|---|---|---|---|---|---|---|---|---|
| 57256 | 20150822 | 02:28-03:17 | 1.2482 | 1.7681 | 33.97 | 256.5 | 20.0 | 39 | 780 | r | |
| 57257 | 20150823 | 02:27-03:08 | 1.2494 | 1.7681 | 33.97 | 256.5 | 20.0 | 36 | 720 | r | |
| 57261 | 20150827 | 02:39-03:22 | 1.2552 | 1.7688 | 33.96 | 256.6 | 60.0 | 16 | 960 | r | |
| 57264 | 20150830 | 02:33-03:18 | 1.2629 | 1.7705 | 33.94 | 256.8 | 20.0 | 35 | 700 | r | |
| 57265 | 20150831 | 02:32-03:23 | 1.2629 | 1.7705 | 33.94 | 256.8 | 60.0 | 25 | 1500 | r | |
| 57275 | 20150910 | 02:58-03:39 | 1.2900 | 1.7779 | 33.82 | 257.9 | 30.0 | 40 | 1200 | r, g, i | 02:58:03:25 for r 26x30s, 03:25-03:30 for g 6x30s, 03:31-03:39 for I 8x30s |
| 57277 | 20150912 | 02:22-03:42 | 1.2967 | 1.7798 | 33.80 | 258.2 | 60.0 | 47 | 2820 | r, g, I | 02:22:03:04 for r 26x60s, 03:10-03:29 for g 13x60s, 03:29-03:42 for i 8x60s |
| 57278 | 20150913 | 02:54-03:38 | 1.3003 | 1.7808 | 33.78 | 258.3 | 60.0 | 30 | 1800 | r | |
| 57283 | 20150918 | 02:54-03:38 | 1.3233 | 1.7868 | 33.70 | 259.2 | 60.0 | 1 | 60 | r | single exposure, bad seeing |
| 57284 | 20150919 | 03:25-03:57 | 1.3233 | 1.7868 | 33.70 | 259.2 | 60.0 | 18 | 1080 | r | |
| 57293 | 20150928 | 03:49-04:13 | 1.3640 | 1.7959 | 33.59 | 260.5 | 60.0 | 16 | 960 | r | |
| 57294 | 20150929 | 03:03-03:48 | 1.3689 | 1.7968 | 33.58 | 260.6 | 60.0 | 30 | 1800 | r | |
| 57295 | 20150930 | 02:41-04:12 | 1.3739 | 1.7978 | 33.57 | 260.8 | 60.0 | 55 | 3300 | r | |
| 57296 | 20151001 | 02:38-04:15 | 1.3790 | 1.7987 | 33.55 | 260.9 | 60.0 | 46 | 2760 | r | |
| 57297 | 20151002 | 02:53-03:35 | 1.3842 | 1.7996 | 33.54 | 261.0 | 60.0 | 19 | 1140 | r | |
| 57305 | 20151010 | 02:53-03:35 | 1.4282 | 1.8054 | 33.47 | 261.9 | 60.0 | 7 | 420 | r | |
| 57315 | 20151020 | 02:53-03:44 | 1.4885 | 1.8087 | 33.40 | 262.4 | 60.0 | 31 | 1860 | r | |
| 57319 | 20151024 | 02:59-04:42 | 1.5141 | 1.8084 | 33.38 | 262.3 | 60.0 | 54 | 3240 | r | |
| 57332 | 20151025 | 02:47-02:49 | 1.5273 | 1.8079 | 33.36 | 262.2 | 60.0 | 1 | 60 | r | single exposure |
| 57321 | 20151026 | 03:14-04:49 | 1.5273 | 1.8079 | 33.36 | 262.2 | 60.0 | 1 | 60 | r | |
| 57325 | 20151030 | 03:52-04:31 | 1.5540 | 1.8060 | 33.32 | 262.0 | 60.0 | 25 | 1500 | r | |
| 57327 | 20151101 | 03:13-04:52 | 1.5674 | 1.8047 | 33.30 | 261.8 | 60.0 | 52 | 3120 | r | |
| 57328 | 20151102 | 03:33-04:53 | 1.5743 | 1.8039 | 33.29 | 261.7 | 60.0 | 48 | 2880 | r | |
| 57329 | 20151103 | 03:28-05:00 | 1.5812 | 1.8031 | 33.28 | 261.5 | 60.0 | 59 | 3540 | r | |
| 57332 | 20151106 | 01:52-02:34 | 1.6014 | 1.8001 | 33.24 | 261.1 | 60.0 | 27 | 1620 | r | |
| 57337 | 20151111 | 03:49-05:10 | 1.6372 | 1.7934 | 33.16 | 260.1 | 60.0 | 52 | 3120 | r | |
| 57338 | 20151112 | 03:48-05:02 | 1.6440 | 1.7919 | 33.14 | 259.9 | 60.0 | 48 | 2880 | r | |
| 57345 | 20151119 | 03:42-04:00 | 1.6874 | 1.7809 | 32.99 | 258.3 | 60.0 | 12 | 720 | r | bad seeing |
| 57350 | 20151124 | 03:18-04:12 | 1.7318 | 1.7670 | 32.77 | 256.3 | 60.0 | 36 | 2160 | r | |
| 57354 | 20151128 | 04:04-04:25 | 1.7617 | 1.7561 | 32.58 | 254.7 | 60.0 | 14 | 840 | r | |
| 57362 | 20151206 | 04:57-05:40 | 1.8148 | 1.7342 | 32.15 | 251.6 | 60.0 | 29 | 1740 | r | |
| 57377 | 20151221 | 01:01-01:57 | 1.9358 | 1.6730 | 30.55 | 242.7 | 60.0 | 34 | 2040 | r | |
| 57381 | 20151225 | 01:46-03.31 | 1.9670 | 1.6556 | 29.97 | 240.1 | 90.0 | 47 | 4230 | r | |
| 57382 | 20151226 | 04:13-05:39 | 1.9753 | 1.6508 | 29.80 | 239.5 | 60.0 | 39 | 2340 | r | |
| 57383 | 20151227 | 02:28-04:07 | 1.9825 | 1.6466 | 29.65 | 238.9 | 90.0 | 48 | 4320 | r | |
| 57385 | 20151229 | 04:01-05:43 | 1.9985 | 1.6374 | 29.29 | 237.5 | 41.0 | 60 | 2460 | r | |
| 57387 | 20151231 | 04:03-03.30 | 2.0140 | 1.6284 | 28.93 | 236.2 | 60.0 | 55 | 3300 | r | |
| 57409 | 20160122 | 02:52-05:13 | 2.1833 | 1.5342 | 23.32 | 222.6 | 90.0 | 45 | 4050 | r | |
| 57410 | 20160123 | 00:43-00:52 | 2.1899 | 1.5311 | 23.04 | 222.1 | 120.0 | 3 | 360 | r | |
| 57411 | 20160124 | 00:55-05:45 | 2.1985 | 1.5271 | 22.66 | 221.5 | 70.0 | 120 | 8400 | r | |
| 57415 | 20160127 | 04:12-05:18 | 2.2295 | 1.5139 | 21.22 | 219.6 | 120.0 | 15 | 1800 | r | no useful data |
| 57417 | 20160130 | 03:43-05:31 | 2.2447 | 1.5081 | 20.48 | 218.8 | 120.0 | 30 | 3600 | r, g, i | 04:27:04:58 for r 10x120s, 03:43-04:13 for g 10x120s, 04:59-05:31 for i 10x120s |
| 57440 | 20160221/22 | 21:41-01:54 | 2.4173 | 1.4910 | 10.48 | 216.3 | 120.0 | 23 | 2760 | r | |
| 57447 | 20160228/29 | 23:47-00:28 | 2.4696 | 1.5097 | 7.27 | 219.0 | 120.0 | 13 | 1560 | r | |
| 57461 | 20160314 | 20:23-22:46 | 2.5794 | 1.5950 | 3.99 | 231.4 | 120.0 | 39 | 4680 | r | |
| 57463 | 20160316 | 19.54-22:04 | 2.5938 | 1.6111 | 4.38 | 233.7 | 120.0 | 39 | 4680 | r | |
| 57479 | 20160401 | 21:39-23:45 | 2.7096 | 1.7820 | 9.72 | 258.5 | 120.0 | 39 | 4680 | r | |
| 57481 | 20160403 | 20:05-20:52 | 2.7231 | 1.8067 | 10.35 | 262.1 | 120.0 | 15 | 1800 | r | |
| 57488 | 20160410/11 | 22:01-00:16 | 2.7734 | 1.9060 | 12.53 | 276.5 | 120.0 | 37 | 4440 | r | |
| 57516 | 20160507/08 | 19:59-00:25 | 2.9598 | 2.3678 | 17.76 | 343.5 | 120.0 | 45 | 5400 | r, g, i | faint coma, 23:32-00:05 for r 15x120s, 22:31-23:224 for g 15x120s, 19:58-20:01 for I 15x120s |
| 57517 | 20160508/09 | 22:33-00:19 | 2.9663 | 2.3862 | 17.86 | 346.1 | 120.0 | 30 | 3600 | r | faint coma |

During the first weeks of the campaign the cometary coma was imaged more frequently – basically during each useful night – in order to monitor possible short-term variability of the comet. Later on and based on the experience on the comet behavior from the initial part of the campaign, the repetition cycle of the nightly images was extended, also to compensate for the decreasing object brightness by longer total integration times of the nightly series. Bias, dark current and sky flat field exposures are available for the observing nights. Since flux calibration of the images is done via tabulated magnitudes of background stars in the individual exposures (see below), no specific and dedicated photometric standard star fields were taken.

The data reduction was performed using a pipeline process developed and tuned at the University Observatory Munich specifically for the Mt. Wendelstein 2m telescope and the WWFI camera. The raw images are subtracted by a monthly master bias in which two features are present: a time-stable vertical line pattern on a 0.02 ADU level and a - on timescales of minutes - varying but uniform off-set level for each port of ~400 ADU. This offset is determined from the three over scan regions of each of the 16 ports and subtracted. The dark current of the detectors is negligible at operational temperature of -115 degrees Celsius. Therefore, usually no dark exposures are subtracted. The images are thereafter divided by a daily master flat which is created from twilight sky flats. The large-scale pattern of the single sky flats. is matched to one sky flat by 4th order polynomials to attain the same illumination before clipping stars and combining the calibrated sky images. Cosmic rays are effectively masked by the edge-detection routine from Gössl et al. (2002). Charge persistence stripes are also masked. The bright regions commencing from saturated, bright stars along the vertical readout direction are contaminated for minutes to hours, depending on the intensity of saturation. The stripes are detected by comparing suspected areas to the local background and are then masked in subsequent exposures. Barely noticeable offsets between the ports remain after the basic data reduction. The offsets are adjusted to an average level by matching the background flux of 30 pixel wide stripes next to the port borders.

Next, the center of the comet was selected by visual inspection. In this step also bad images, where the comet was contaminated by dead or masked columns, were removed. For the final images the comet core was centered to a 1601 x 1601 pixel grid matching a size of 5.34 x 5.34 arcmin on the sky. In the last step the images were photometrically aligned by their zero point and the sky was subtracted.

The determination of the photometric zero point in the individual images was adjusted to the Pan-STARRS-3 Pi source catalog (PV3) in AB magnitudes (Magnier et al. 2013), which filter system in turn agrees well with SDSS AB magnitudes. Tonry et al. (2012) determined the conversions between the PS1 and SDSS filter system and indicates a small color correction between PS1 and SDSS of 0.007*(g-r). Kosyra et al. (2014) measured the color corrections between WWFI and SDSS to 0.035*(r-i). Tab. 2 gives an overview of the bandwidth specifications of the filters used.

The differences for the r band are only marginal, i.e the r SDSS filter is shifted in $\lambda_{eff}$ about 4 nm to the red (at the flanks shifted less than 20 nm into the red) which may explain small differences for the color terms. Note in particular that all photometric data is reported in the SDSS system.

**Table 2**  Passband parameters of the g, r, i filters for WWFI, PS1, and SDSS

|  | $\lambda_{eff}$ [nm] | $\lambda_{blue}$ [nm] | $\lambda_{red}$ [nm] |
|---|---|---|---|
| g band |  |  |  |
| WWFI | 480.3 | 418 | 550 |
| PS1 | 483.2 | 414 | 552 |
| SDSS | 476.6 | 411 | 561 |
| r band |  |  |  |
| WWFI | 619.3 | 557 | 695 |
| PS1 | 618.8 | 548 | 694 |
| SDSS | 622.7 | 563 | 713 |
| i band |  |  |  |
| WWFI | 758.6 | 698 | 836 |
| PS1 | 752.5 | 688 | 824 |
| SDSS | 764.2 | 699 | 869 |

**Figure 1**  Example images from the structure enhancement processing of 67P from night 10 September 2015. The images show: Left column – Top panel: the full frame composite of the comet (290000 x 290000 km); Bottom panel: Isophote pattern of the cometary coma (65000 x 65000 km). Middle column – Top panel: Laplace filtered image of the coma, filter with 15 pixels (65000 x 65000 km); Middle panel: Laplace filtered image of the coma, filter width 23 pixels (65000 x 65000 km); Bottom panel: Laplace filtered image of the coma, filter width 47 pixels (130000 x 130000 km); Right column, Top panel: Renormalized image, ratio of original by average coma profile (65000 x 65000 km); Bottom panel: Renormalized image, difference between original and average coma profile (65000 x 65000 km). The listed distances describe the field of view at the distance of the comet. Image orientation is North up, East to the left.

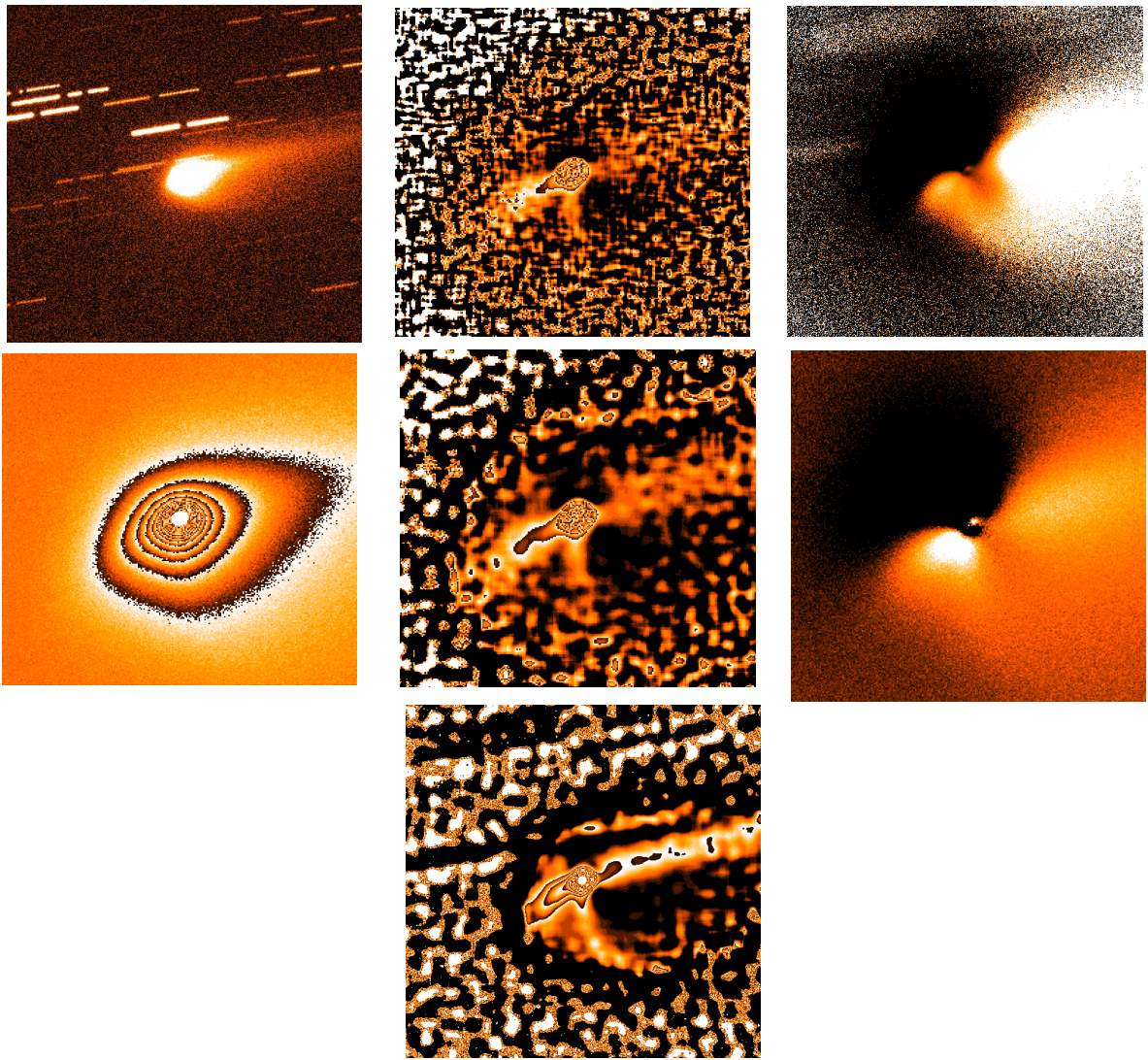

## 3. Data Post-Processing for the Analysis of the Cometary Images

From the flux calibrated image series of a single night the mean and the median-averaged composite frame is produced after proper centering of the individual frames to the central brightness peak in the coma. The typical centering accuracy is usually better than 0.2 pixels, i.e. below about 45 to 70 km at the distance of the comet. While the mean coma image contains all signatures from background objects as well as in parts from artificial and natural image artifacts, the median-averaged composite of a night (after photometric and sky alignment) is smoothed and widely cleaned from high signal levels of the background objects and image artifacts. However, it is noted that remnant signatures at the level of the local coma intensity level remain in the median composite images. Usually, no disturbing impact results from these contamination for the further data analysis of the comet images (see below). The averaged composite image of a single night provides the summary of the coma appearance of the nightly observing window, but it does not allow to evaluate the temporal evolution of the coma during the nightly observing interval. Any such assessment requires analysis of individual images of the night or subsets of them. The average composite images were removed from a minor sky background level remaining from the averaging process of the individual sky-removed images.

For the assessment of the cometary activity further analysis methods are applied. The search for coma and tail structures starts with a visual inspection of the isophote pattern of the coma and tail for time variable and asymmetric patterns deviating from 'tear-drop-shape' coma isophotes. Furthermore, two structure enhancement methods are applied: (1) Adaptive Laplace filtering (Boehnhardt&Birkle 1994) using different filtering width for narrow (7 pixels wide) to extended (128 pixels wide) features, and (2) coma renormalization (A'Hearn et al. 1986) by either dividing or subtracting a two-dimensional (2D) coma profile created from the mean radial coma profile obtained from the comet image. For the latter the original and the 2D-profile image are aligned to each other via the central brightness peaks in the coma. While the adaptive Laplace filtering is very sensitive to low-level systematic local and global brightness differences in the coma and tail region, it reduces the effective spatial resolution of the noticeable structures in the image according to the filter width chosen. Moreover, in our application it introduces a non-linear scaling of the enhanced structure such that only geometric, but no quantitative photometric conclusions can be obtained. The renormalization method is less sensitive to low level structures, but keeps the linear intensity scaling of the result image as well as it does not reduce the geometric resolution of the result frame. However, artifacts (artificial structures) can result from decentering between the original and the 2D-profile image, in particular close to the coma center. Coma and tail structures found by either one of the enhancement methods described above, need to be verified, ideally, by the respective alternative method and in any case by an inspection of the original image for systematic irregularities in its isophote pattern. The two enhancement methods are applied to all individual exposures of the comet as well as to the mean and median averaged composite frames of the individual nights. Inspections of the time series of processed individual images are used for an assessment of short-term variability of coma structures identified. The mean and median averaged composites provide a much higher signal-to-noise for an extended and quantitative assessment of the coma, in particular also for the extended geometry of coma features. Fig. 1 shows an example of a composite image of 67P processed by the enhancement methods described above.

For the assessment of the global coma status radial profiles of the median averaged composite images of the individual nights are considered. The radial profiles are obtained by flux integration in concentric rings of increasing radius centered on the central brightness peak in the coma. The gradient slope in the average radial profile is approximated in a linear fit to the log-log pairs for the flux versus the distance in the coma. Different distance ranges, 5000 – 10000 km, 10000 – 15000 km, 15000 – 20000 km, and 20000 – 25000 km are used for the slope fitting. The different ranges allow for the assessment of possible changes in the slope parameter with radial distance from the nucleus, originating from, for instance, radiation pressure effects on the coma dust, from gas contamination in the broadband filter images, and from physical changes of the coma content.

## 4. Results

Comet 67P displayed an extended coma over the whole observing interval from 22 August 2015 (1.24 AU post-perihelion) until the end of the observations on 9 May 2016 (2.97 AU post-perihelion). A direct signature of the cometary nucleus is not detected. It is assumed to be hidden in the central brightness peak of the coma produced by the atmospheric seeing. Thus, the nucleus may still contribute to the total brightness of the coma measured in circular apertures centered on the central brightness peak. The available images and measured data allow to characterize the cometary activity and properties of the coma, i.e. the magnitude, colors and spectral slopes, radial coma profiles, dust activity, coma structures and tail appearance.

*Coma magnitudes:* In the r filter images the coma magnitude was measured centered on the nucleus position for 4 aperture radii, i.e. for 5000, 10000, 15000 and 20000 km at the distance of the comet (Fig. 2, Tab. 3). Maximum brightness of the comet is found for the period of late August to early September 2015. Since the observations start only 9 days after perihelion, the period of maximum brightness is not well constraint from our data. Between about late September and at least to the end of December 2015 the coma magnitude followed a decrease, approximately linearly with time. As of the last third of January 2016 the comet displayed a close to constant coma magnitude in the measured apertures until the end of February 2016 when it started to get fainter again. Typical measurement errors are in the range 0.03-0.05mag, except for a few nights with bad seeing and when the coma got very faint in May 2016 (errors ~0.1 mag). Variable coma brightness during the

**Figure 2**  Coma magnitude of 67P/Churyumov-Gerasimenko, measured in the WWFI r filter, versus time. The results for 4 different projected aperture radii at the distance of the comet (5000, 10000, 15000, 20000 km) are plotted. The time is give as Modified Julian Date MJD).

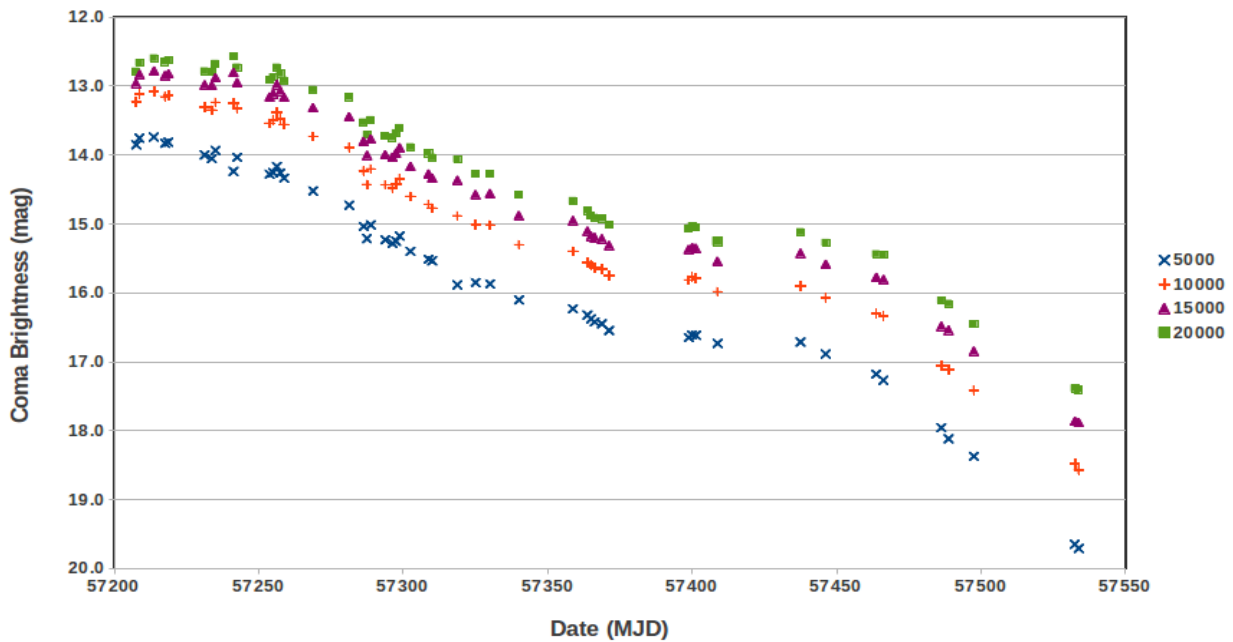

**Table 3** Radial gradient, magnitude and afρ parameter of the dust coma of 67P during the observing period. The table lists the Modified Julian Date (MJD) and UT date of the observation, the Sun (r in AU) and Earth (Δ in AU) distance of the comet, its phase angle (deg) and the slope, magnitude (mag) and afρ parameter of the dust coma. The latter three quantities are listed for 4 distance range (5000/10000, 10000/15000, 15000/20000, 20000/25000 km) and 4 aperture radii (5000, 10000, 15000, 20000 km). NM means not measured or calculated. The uncertainty of magnitude measurements is mostly about 0.03 to 0.05mag, that of the radial slope is about 0.02.

| 67P | Dust Coma | | | | Slope | Slope | Slope | Slope | Magnitude | Magnitude | Magnitude | Magnitude | Afρ (cm) | Afρ (cm) | Afρ (cm) | Afρ (cm) |
|---|---|---|---|---|---|---|---|---|---|---|---|---|---|---|---|---|
| MJD | Date (UT) | r (AU) | Δ (AU) | Phase (deg) | 5000/10000 km | 10000/15000 km | 15000/20000 km | 20000/25000 km | 5000 km | 10000 km | 15000 km | 20000 km | 5000 km | 10000 km | 15000 km | 20000 km |
| 57256 | 20150822 | 1.2482 | 1.7681 | 33.97 | 0.79 | 0.62 | 0.52 | 0.46 | 13.85 | 13.23 | 12.96 | 12.79 | 369 | 327 | 280 | 246 |
| 57257 | 20150823 | 1.2494 | 1.7681 | 33.97 | 0.82 | 0.64 | 0.51 | 0.44 | 13.76 | 13.11 | 12.83 | 12.67 | 404 | 366 | 316 | 276 |
| 57261 | 20150827 | 1.2552 | 1.7688 | 33.96 | 0.84 | 0.68 | 0.57 | 0.50 | 13.74 | 13.08 | 12.78 | 12.60 | 415 | 383 | 334 | 297 |
| 57264 | 20150830 | 1.2629 | 1.7705 | 33.94 | 0.86 | 0.73 | 0.65 | 0.62 | 13.83 | 13.16 | 12.85 | 12.65 | 387 | 359 | 318 | 287 |
| 57265 | 20150831 | 1.2629 | 1.7705 | 33.94 | 0.87 | 0.71 | 0.62 | 0.56 | 13.82 | 13.13 | 12.83 | 12.63 | 393 | 368 | 325 | 293 |
| 57275 | 20150910 | 1.2900 | 1.7779 | 33.82 | 1.03 | 1.01 | 0.96 | 0.93 | 14.00 | 13.31 | 12.98 | 12.79 | 348 | 330 | 296 | 266 |
| 57277 | 20150912 | 1.2967 | 1.7798 | 33.80 | 0.93 | 0.77 | 0.65 | 0.56 | 14.05 | 13.35 | 12.99 | 12.79 | 336 | 321 | 297 | 268 |
| 57278 | 20150913 | 1.3003 | 1.7808 | 33.78 | 0.92 | 0.77 | 0.66 | 0.58 | 13.93 | 13.24 | 12.88 | 12.68 | 377 | 358 | 331 | 300 |
| 57283 | 20150918 | 1.3233 | 1.7868 | 33.70 | 1.30 | 0.94 | 0.76 | 0.64 | 14.24 | 13.25 | 12.81 | 12.57 | 295 | 368 | 368 | 343 |
| 57284 | 20150919 | 1.3233 | 1.7868 | 33.70 | 0.93 | 0.81 | 0.78 | 0.71 | 14.04 | 13.33 | 12.95 | 12.73 | 358 | 344 | 325 | 298 |
| 57293 | 20150928 | 1.3640 | 1.7959 | 33.59 | 0.95 | 0.84 | 0.74 | 0.68 | 14.28 | 13.54 | 13.16 | 12.91 | 308 | 303 | 287 | 271 |
| 57294 | 20150929 | 1.3689 | 1.7968 | 33.58 | 1.01 | 0.89 | 0.78 | 0.71 | 14.26 | 13.50 | 13.11 | 12.87 | 317 | 318 | 303 | 283 |
| 57295 | 20150930 | 1.3739 | 1.7978 | 33.57 | 1.03 | 0.89 | 0.77 | 0.68 | 14.17 | 13.38 | 12.98 | 12.73 | 347 | 359 | 345 | 325 |
| 57296 | 20151001 | 1.3790 | 1.7987 | 33.55 | 1.03 | 0.89 | 0.79 | 0.71 | 14.26 | 13.47 | 13.07 | 12.82 | 321 | 331 | 319 | 301 |
| 57297 | 20151002 | 1.3842 | 1.7996 | 33.54 | 1.02 | 0.89 | 0.78 | 0.68 | 14.34 | 13.56 | 13.17 | 12.92 | 302 | 308 | 295 | 277 |
| 57305 | 20151010 | 1.4282 | 1.8054 | 33.47 | 1.06 | 0.98 | 0.89 | 0.84 | 14.52 | 13.73 | 13.32 | 13.06 | 272 | 281 | 275 | 263 |
| 57315 | 20151020 | 1.4885 | 1.8087 | 33.40 | 1.11 | 1.01 | 0.91 | 0.81 | 14.73 | 13.89 | 13.45 | 13.16 | 245 | 266 | 266 | 260 |
| 57319 | 20151024 | 1.5141 | 1.8084 | 33.38 | 1.06 | 0.99 | 0.88 | 0.77 | 15.04 | 14.24 | 13.80 | 13.53 | 191 | 200 | 198 | 192 |
| 57332 | 20151025 | 1.5273 | 1.8079 | 33.36 | 1.05 | 0.99 | 0.94 | 0.84 | 15.21 | 14.43 | 14.01 | 13.71 | 164 | 168 | 166 | 164 |
| 57321 | 20151026 | 1.5273 | 1.8079 | 33.36 | 1.07 | 0.99 | 0.89 | 0.77 | 15.01 | 14.21 | 13.77 | 13.49 | 198 | 209 | 208 | 201 |
| 57325 | 20151030 | 1.5540 | 1.8060 | 33.32 | 1.06 | 0.99 | 0.88 | 0.77 | 15.23 | 14.44 | 14.00 | 13.72 | 168 | 175 | 174 | 168 |
| 57327 | 20151101 | 1.5674 | 1.8047 | 33.30 | 1.07 | 1.00 | 0.90 | 0.79 | 15.28 | 14.48 | 14.03 | 13.75 | 163 | 171 | 171 | 167 |
| 57328 | 20151102 | 1.5743 | 1.8039 | 33.29 | 1.10 | 1.01 | 0.91 | 0.80 | 15.25 | 14.42 | 13.97 | 13.69 | 168 | 181 | 183 | 178 |
| 57329 | 20151103 | 1.5812 | 1.8031 | 33.28 | 1.11 | 1.01 | 0.91 | 0.80 | 15.18 | 14.34 | 13.90 | 13.61 | 182 | 196 | 198 | 193 |
| 57332 | 20151106 | 1.6014 | 1.8001 | 33.24 | 1.06 | 1.01 | 0.92 | 0.81 | 15.40 | 14.60 | 14.17 | 13.89 | 152 | 158 | 157 | 153 |
| 57337 | 20151111 | 1.6372 | 1.7934 | 33.16 | 1.06 | 1.01 | 0.92 | 0.81 | 15.52 | 14.72 | 14.28 | 13.98 | 141 | 147 | 148 | 146 |
| 57338 | 20151112 | 1.6440 | 1.7919 | 33.14 | 1.02 | 0.98 | 0.90 | 0.80 | 15.54 | 14.77 | 14.34 | 14.04 | 140 | 141 | 141 | 139 |
| 57345 | 20151119 | 1.6874 | 1.7809 | 32.99 | 1.32 | 1.11 | 0.98 | 0.89 | 15.88 | 14.88 | 14.37 | 14.07 | 106 | 132 | 142 | 141 |
| 57350 | 20151124 | 1.7318 | 1.7670 | 32.77 | 1.07 | 1.02 | 0.93 | 0.84 | 15.85 | 15.01 | 14.58 | 14.28 | 113 | 122 | 122 | 120 |
| 57354 | 20151128 | 1.7617 | 1.7561 | 32.58 | 1.09 | 1.01 | 0.93 | 0.84 | 15.87 | 15.02 | 14.56 | 14.28 | 113 | 124 | 126 | 123 |
| 57362 | 20151206 | 1.8148 | 1.7342 | 32.15 | 1.06 | 1.01 | 0.95 | 0.85 | 16.10 | 15.31 | 14.88 | 14.58 | 94 | 98 | 97 | 96 |
| 57377 | 20151221 | 1.9358 | 1.6730 | 30.55 | 1.06 | 0.99 | 0.93 | 0.87 | 16.23 | 15.40 | 14.95 | 14.67 | 89 | 96 | 96 | 94 |
| 57381 | 20151225 | 1.9670 | 1.6556 | 29.97 | 1.05 | 1.00 | 0.93 | 0.85 | 16.32 | 15.56 | 15.11 | 14.81 | 83 | 84 | 85 | 83 |
| 57382 | 20151226 | 1.9753 | 1.6508 | 29.80 | 1.04 | 0.99 | 0.93 | 0.88 | 16.38 | 15.60 | 15.18 | 14.88 | 79 | 81 | 79 | 78 |
| 57383 | 20151227 | 1.9825 | 1.6466 | 29.65 | 1.05 | 1.00 | 0.95 | 0.89 | 16.42 | 15.64 | 15.20 | 14.91 | 76 | 78 | 78 | 76 |
| 57385 | 20151229 | 1.9985 | 1.6374 | 29.29 | 1.05 | 1.00 | 0.93 | 0.83 | 16.45 | 15.66 | 15.22 | 14.92 | 74 | 77 | 77 | 76 |
| 57387 | 20151231 | 2.0140 | 1.6284 | 28.93 | 1.06 | 1.00 | 0.94 | 0.88 | 16.55 | 15.75 | 15.31 | 15.02 | 68 | 71 | 71 | 70 |
| 57409 | 20160122 | 2.1833 | 1.5342 | 23.32 | 1.07 | 1.03 | 0.94 | 0.90 | 16.65 | 15.82 | 15.37 | 15.07 | 65 | 70 | 70 | 69 |
| 57410 | 20160123 | 2.1899 | 1.5311 | 23.04 | 1.05 | 0.99 | 0.91 | 0.81 | 16.61 | 15.77 | 15.35 | 15.04 | 67 | 73 | 72 | 72 |
| 57411 | 20160124 | 2.1985 | 1.5271 | 22.66 | 1.07 | 1.00 | 0.94 | 0.88 | 16.62 | 15.79 | 15.35 | 15.05 | 67 | 72 | 72 | 71 |
| 57415 | 20160127 | 2.2295 | 1.5139 | 21.22 | NM | NM | NM | NM | NM | NM | NM | NM | NM | NM | NM | NM |
| 57417 | 20160130 | 2.2447 | 1.5081 | 20.48 | 1.03 | 0.99 | 0.94 | 0.91 | 16.73 | 15.99 | 15.54 | 15.26 | 61 | 61 | 61 | 60 |
| 57440 | 20160221/22 | 2.4173 | 1.4910 | 10.48 | 1.08 | 1.08 | 1.07 | 1.04 | 16.71 | 15.90 | 15.43 | 15.12 | 71 | 75 | 77 | 77 |
| 57447 | 20160228/29 | 2.4696 | 1.5097 | 7.27 | 1.12 | 1.08 | 1.02 | 0.97 | 16.89 | 16.07 | 15.59 | 15.28 | 64 | 68 | 71 | 71 |
| 57461 | 20160314 | 2.5794 | 1.5950 | 3.99 | 1.14 | 1.16 | 1.09 | 1.03 | 17.18 | 16.30 | 15.78 | 15.44 | 60 | 68 | 73 | 74 |
| 57463 | 20160316 | 2.5938 | 1.6111 | 4.38 | 1.18 | 1.19 | 1.13 | 1.05 | 17.27 | 16.34 | 15.81 | 15.45 | 57 | 67 | 73 | 76 |
| 57479 | 20160401 | 2.7096 | 1.7820 | 9.72 | 1.21 | 1.25 | 1.20 | 1.12 | 17.96 | 17.06 | 16.48 | 16.11 | 40 | 46 | 52 | 55 |
| 57481 | 20160403 | 2.7231 | 1.8067 | 10.35 | 1.36 | 1.28 | 1.25 | 1.30 | 18.12 | 17.11 | 16.54 | 16.16 | 36 | 46 | 51 | 55 |
| 57488 | 20160410/11 | 2.7734 | 1.9060 | 12.53 | 1.26 | 1.28 | 1.25 | 1.17 | 18.37 | 17.41 | 16.85 | 16.45 | 33 | 40 | 45 | 48 |
| 57516 | 20160507/08 | 2.9598 | 2.3678 | 17.76 | 1.45 | 1.44 | 1.45 | 1.33 | 19.65 | 18.48 | 17.86 | 17.39 | 18 | 26 | 31 | 36 |
| 57517 | 20160508/09 | 2.9663 | 2.3862 | 17.86 | 1.45 | 1.66 | 1.49 | 1.31 | 19.71 | 18.57 | 17.88 | 17.41 | 17 | 25 | 31 | 36 |

nightly exposure window beyond the uncertainty range for the individual nights is not found for the 4 aperture sizes used. Variations over several nights are generally slow as well as of small amplitude and they appear to be associated with the long-term activity evolution of the comet. Brightness enhancements beyond 0.2mag, indicating a short-term outburst of the comet, are not identified in our observations.

*Coma colors:* The color of the coma dust is calculated from the flux calibrated images of 67P and is obtained for 4 observing epoch, each using 3 pairs of filters (g-r, r-i and g-i), i.e. for 10 and 12 September 2015, for 30 January 2016, and for 7/8 May 2016 (see Tab. 4). Using the central wavelengths of the filters as a reference, the equivalent average spectral slope S can be estimated for the wavelength range covered by the filters (see Tab. 2) using formula (1)

(1) $$S = [10^{(-0.4 * (C_{comet} - C_{sun}))}] / [\lambda_{long} - \lambda_{short}]$$

with $C_{comet}$ and $C_{sun}$ as the magnitude difference of the comet and Sun, respectively, measured for the filter pair with effective central wavelengths $\lambda_{long}$ and $\lambda_{short}$.

**Table 4** Filter colors and average spectral slope of the 67P coma for different aperture sizes. Filter colors and spectral slopes are given for the WWFI g, r, i filters and for 4 coma aperture radii (5000, 10000, 15000, 20000 km) as well as the mean values of the results for the aperture radii. Solar colors used are 0.44 mag for g-r, 0.11 mag for r-i and 0.55 mag for g-i. The larger scatter of the results for 7/8 May 2016 is due to higher uncertainties in the coma photometry because of the very much reduced cometary brightness.

| Filter Range | g-r | g-r | g-r | g-r | g-r | r-i | r-i | r-i | r-i | r-i | g-i | g-i | g-i | g-i | g-i | |
|---|---|---|---|---|---|---|---|---|---|---|---|---|---|---|---|---|
| Coma Radius (km) | 5000 | 10000 | 15000 | 20000 | mean | 5000 | 10000 | 15000 | 20000 | mean | 5000 | 10000 | 15000 | 20000 | mean | |
| 20150910 | 0.59 | 0.56 | 0.53 | 0.50 | **0.55** | 0.22 | 0.22 | 0.21 | 0.21 | **0.22** | 0.81 | 0.78 | 0.75 | 0.71 | **0.76** | Colour |
|  | 10.5 | 8.4 | 6.4 | 4.3 | **7.4** | 7.9 | 7.7 | 7.1 | 6.7 | **7.4** | 9.8 | 8.5 | 7.1 | 5.7 | **7.8** | Spectral Gradient |
| 20150912 | 0.58 | 0.57 | 0.54 | 0.51 | **0.55** | 0.25 | 0.23 | 0.23 | 0.22 | **0.23** | 0.83 | 0.80 | 0.76 | 0.73 | **0.78** | Colour |
|  | 9.9 | 8.8 | 6.7 | 4.6 | **7.5** | 10.2 | 8.5 | 8.0 | 7.6 | **8.6** | 10.7 | 9.2 | 7.7 | 6.3 | **8.5** | Spectral Gradient |
| 20160130 | 0.62 | 0.61 | 0.61 | 0.61 | **0.61** | 0.22 | 0.23 | 0.24 | 0.25 | **0.23** | 0.84 | 0.84 | 0.85 | 0.86 | **0.85** | Colour |
|  | 12.9 | 12.6 | 12.4 | 12.2 | **12.5** | 7.7 | 8.2 | 9.0 | 9.6 | **8.6** | 11.0 | 11.1 | 11.5 | 11.7 | **11.3** | Spectral Gradient |
| 20160507 | 0.48 | 0.62 | 0.64 | 0.67 | **0.60** | 0.35 | 0.24 | 0.18 | 0.14 | **0.23** | 0.82 | 0.85 | 0.82 | 0.82 | **0.83** | Colour |
|  | 2.4 | 12.7 | 14.5 | 17.2 | **11.7** | 17.5 | 8.9 | 4.6 | 2.2 | **8.3** | 10.3 | 11.6 | 10.0 | 10.0 | **10.5** | Spectral Gradient |

Due to the larger solar distance (2.24 and 2.96 AU), the gas emission in the 67P coma in the g, r, i filter ranges should be negligibly low on 30 January and 7/8 May 2016, such that the measured colors and estimated spectral slopes can be considered to represent the values of the cometary dust. At that time (phase angle of the comet was 20.5 deg and 17.8 deg, respectively) the reddening of the dust coma was about 12 %/100nm in g-r wavelength range, about 8.5 %/100nm for r-i, and about 11 %/100nm for g-i. The dust reddening of 67P is slightly decreasing with increasing wavelength, a phenomenon seen for instance for the surface of 67P (Fornasier et al. 2015), for other cometary nuclei (Lamy et al. 2004) and for Kuiper Belt objects (Doressoundiram et al. 2007). Within the measurement uncertainties these values also agree with the colors and spectral gradients determined for the nucleus of 67P by Tubiana et al. (2008, 2011) and Lowry (2006). It is slightly below the spectral slope determined from in-situ measurements of dust grains in the coma of 67P (Cremonese et al. 2015). No clear increase of the reddening with phase angle is seen in our data. Although at the detection limit, there might be increasing and decreasing trends in the coma colors and spectral slopes with increasing distance in the coma. In two-dimensional color images no reddening variations could be identified that can be associated to coma structures (described below).

The mean g-r and g-i colors and spectral slopes are lower on 10 and 12 September 2015 than in January and May 2016 while the r-i colors seem to agree within the errors for all four observing epochs. This is taken as indication that gas emission, i.e. mostly from $C_2$ molecules, is contributing significantly to the measured coma flux in the g filter on 10 and 12 September 2015, when the comet had passed perihelion for about a month (solar distance of about 1.3 AU). The agreement of the r-i colors and spectral slopes for all 4 observing epochs with the comet at 1.3, 2.2, and 2.9 AU, respectively, is seen as an argument to assume that the r-i wavelength region is not significantly contaminated by gas emissions and represents thus the comet appearance in dust reflected sunlight. Possible weak gas contamination in this spectral range could come from $C_2$, $NH_2$ molecules and from the [OI] line in the coma, but is usually not noticeable in images of comets taken with broadband filters in the red wavelength region.

*Dust activity:* The dust activity of comets is typically characterized by determining the afρ parameter of the coma in a wavelength region where dust reflected sunlight dominates the coma brightness. The afρ approach for the characterization of the dust activity of comets was introduced by A'Hearn et al. (1984) and provides a quantitative equation that connects the projected aperture radius ρ at the distance of the comet, the so called filling factor f of the cometary dust in the light path and the albedo a of the dust (which is assumed to be constant for all reflecting dust grains) with the measured coma flux.

(2)        $af\rho = 4 * \Delta^2 * r^2 / \rho * f_{comet} / f_{sun}$

In eq. (2) $\Delta$ and r denote the Earth and Sun distance of the comet, respectively, and $f_{comet}$ and $f_{sun}$ are the fluxes of the comet and Sun, respectively, in the filter band used for the measurements. Note that for a 'quasi-steady-state' coma with homogeneous and isotropic dust expansion, constant production level of the nucleus for the dust in the measurement aperture, and neglecting changes of the dust properties due to radiation pressure effects along their travel through the measurement aperture, the afρ parameter should be independent from the size of the measurement aperture.

**Figure 3** Estimated percentage of nucleus brightness to the total cometary brightness for different measurement aperture radii (5000, 10000, 15000, 20000 km). The ratio is plotted over the full period of the Mt. Wendelstein observations of the comet.

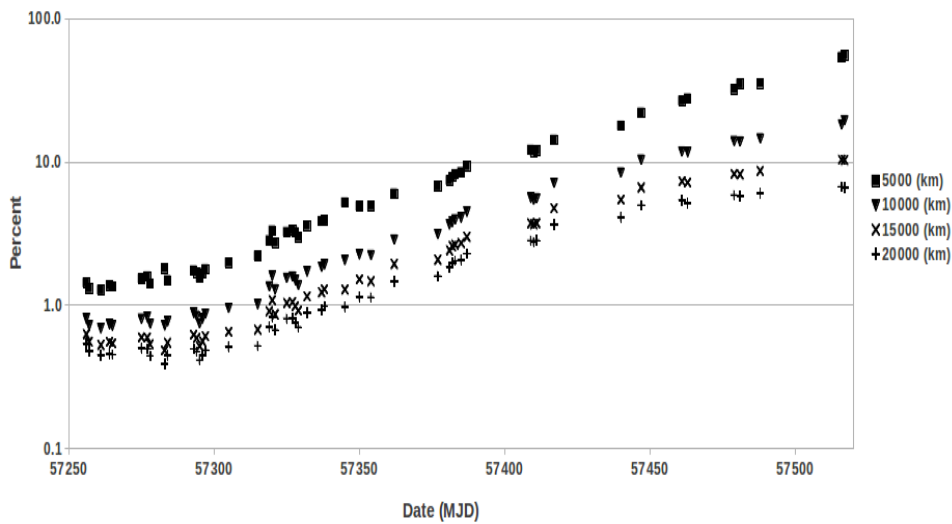

For the estimation of the light contribution of the 67P nucleus to the total brightness measured in circular apertures centered on the central brightness peak in the coma, we applied the standard formulas (equations (3) and (5) in Lamy et al. 2004) and the values for the equivalent radius (1.98 km; Lamy et al. 2008), albedo (0.059; Sierks et al. 2015) and linear phase darkening (0.04 mag/deg; Snodgrass et al. 2013). The parameters used for the nucleus light calculations are to be seen as approximation only, since an accurate prediction of the nucleus brightness (considering shape, rotation, surface reflectivity and light scattering geometry) is not available. Fig. 3 shows the results for different aperture radii: For the largest aperture used (20000 km radius) the nucleus contribution is below 1 percent until mid Nov. 2015 and does not exceed 7 percent by early May 2016. The smallest aperture (5000 km radius) receives less than 2 percent of nucleus light until about mid October 2015, but has a high contribution of more than 50 percent towards the end of the Wendelstein observations of 67P. For the 10000km radius aperture – which is considered a good reference for comparison with other publications – the level of 5 percent nucleus light contribution was reached in January 2016.

As argued from the color measurements (see above) for 67P, the best – since widely free of gas contamination – wavelength region for assessing the dust activity are in the WWFI r and i filter pass bands. Because of the large number of data available, we thus use the r filter images for determining the af$\rho$ values of 67P over time, again applying aperture radii of 5000, 10000, 15000, and 20000 km at the distance of the comet (see Tab. 3). Fig. 4 shows the evolution of af$\rho$ versus time, derived directly from the WWFI r filter brightnesses of the comet.

After a plateau during late August 2015 which represents the likely maximum activity level mentioned above, the dust activity of 67P decreased almost linearly with time until at least the end of December 2015. As of late January 2016 a bump is noticeable in the af$\rho$ evolution lasting until at least early April 2016. During late September 2015 until January 2016 the af$\rho$ curves for different aperture radii in the cometary coma show only minor differences. This indicates that the dust coma was in 'quasi-steady-state' very close to the assumption made for the af$\rho$ assessment via eq. (3). During this time interval a small trend of decreasing af$\rho$ values with increasing aperture size during a single measurement epoch is seen for 67P. During late August until mid September 2015 and as of February 2016 the af$\rho$ values for the 4 aperture radii spread further apart in a systematic manner, i.e. during the earlier period the af$\rho$ values for smaller aperture radii are higher and vice versa during the latter time period. This indicates that the dust coma of the comet was not in 'quasi-steady-state' and is discussed further below.

Fig. 4 suggests that the dust production represented by the af$\rho$ measurements follows approximately an exponential law with solar distance r, i.e. afr ~ $r^\gamma$. The radial exponent $\gamma$ is obtained by a least square fit to the af$\rho$ data for the time interval of the 'quasi-steady-state' dust coma, i.e. from 27 September to 30 December 2015. The results for the heliocentric exponent of the dust activity decrease of 67P is listed in Tab. 5 for the 10000 km radius reference aperture. Other aperture radii (5000, 15000, 20000 km) give very similar results within the uncertainties of the fits.

In order to assess the impact of the brightness enhancement due to phase function effects in the light scattering of the coma dust, we apply different phase function corrections to the measured coma brightness of 67P for the 10000km aperture radius. In general, the phase function reduces the reflected coma magnitude value. For small phase angles $\alpha$ (typically up to 30 deg) a linear reduction with $\alpha$ can be applied with typical parameters $\beta$ for comets between 0.02 and 0.06 (Lamy et al. 2004). Alternatively, the two-parameter phase function of Schleicher et al. (1998), based on 1P/Halley measurements, is available or the phase angle correction by Müller (1999) which also considers geometric projection effects by a non-spherical coma, symmetric to the radial direction with respect to the Sun. Fornasier et al. (2015) obtained results for the 67P nucleus surface reflectivity applying a HG-type phase function (Bowell et al. 1989) with a best fit G of -0.13. Fig. 5 shows the phase angle corrected af$\rho$ values of 67P for various phase function solutions. In Fig. 5 $\beta$ = 0.04 is used as mean value for the linear phase function which is lower than the $\beta$ value found for the nucleus of 67P (0.059 – 0.076; Tubiana et al. 2011; Lowry et al. 2012). A second HG-type phase function is applied as well with G = 0.15 that is found from many asteroid light curves (Tedesco 1989). Obviously, the af$\rho$ value for zero phase angle depends on the phase darkening and is highest for the Bowell-type phase function with G = -0.13 (as obtained for the 67P nucleus by Fornasier et al. 2015). Two phase functions provide a close to linear decrease of af$\rho$ with solar distance over the time interval from the end of September 2015 to beginning of April 2016, i.e. the linear phase function with $\beta$ = 0.04 and the Bowell-HG-type phase function with G = 0.15. The phase functions of Schleicher et al. (1998) and of Müller (1999) can approximate the af$\rho$ data well between end September and end December 2015, but do not describe well the 67P dust activity measured in 2016.

**Figure 4** Evolution of the dust activity of 67P: af$\rho$ versus solar distance (AU). The af$\rho$ parameter, determined from r filter images, is plotted for 4 measurement aperture radii (5000, 10000, 15000, 20000 km). Y axis (af$\rho$) is logarithmic, X axis (solar distance) is linear. The broken lines in the lower panel represent the best fit results for the observed af$\rho$ versus solar distance for the 4 aperture radii. The fit is performed for measurements from 27 September to 30 December 2015.

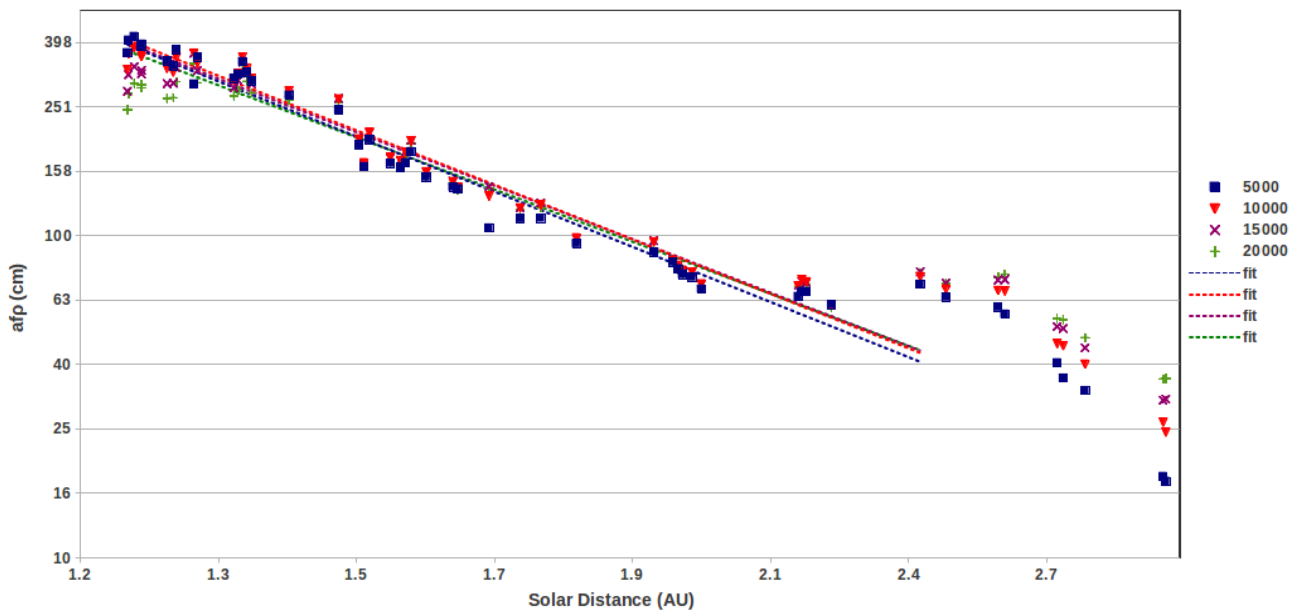

The dust activity of 67P after mid September 2015 until early May 2016 can be described as an exponential decrease by adopting either a linear phase function with $\beta$ ~ 0.04 or a HG-type phase function with G ~ 0.15. Phase function corrections after Schleicher and Müller require deviating dust activity around the time of opposition passage of 67P (January to April 2016), while the HG-type phase corrections of Fornasier et al. results in a small drop of the dust activity around that time period. Based on our observation a continuous exponential decrease of af$\rho$ with increasing heliocentric distance appears to be likely, since there is no clear evidence for enhanced activity in the light curve for the 2016 opposition passage of 67P.

**Figure 5**   Dust parameter afρ of 67P versus solar distance, corrected for different phase functions. The dashed lines indicate best fit solutions for the power law of the various phase-function-corrected afρ curves versus solar distance. All afρ values refer to coma brightness measurements of 67P through a 10000 km radius aperture. Symbols and dashed lines represent the measured and best fit values of afρ for the respective phase functions corrected data.

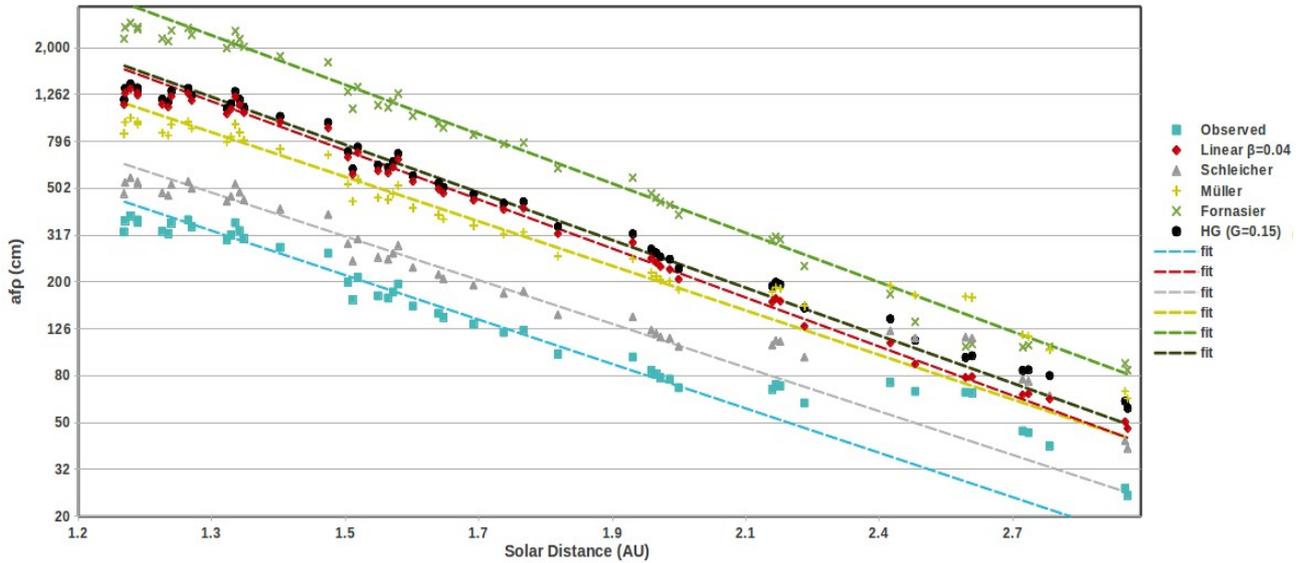

**Table 5**   Heliocentric distance exponent for the afρ decrease of the coma of 67P. The table lists for various phase function models the best fit exponent γ to the afρ data in a 10000 km radius aperture for the time interval 27 September to 30 December 2015. Phase function used are 'none', 'linear' , 'Schleicher', 'Müller', 'Fornasier' and 'Bowell' as described in the text.

| Phase Correction | Dust Activity Exponent |
|---|---|
| None | -3.79 +/-0.13 |
| Linear | -4.19 +/-0.12 |
| Schleicher | -3.73 +/-0.14 |
| Müller | -3.80 +/-0.13 |
| Fornasier | -4.21 +/-0.12 |
| Bowell | -4.07 +/-0.13 |

The exponent for the heliocentric activity decrease fitted to our observations from 27 September to 30 December 2015 are listed in Tab. 5. They vary depending on the adopted phase function approximation between about -3.7 and -4.2. Considering nucleus brightness contributions in the aperture measurements (not done in Tab. 5 and in the discussion below) will change the heliocentric afρ exponent by -0.1. Different authors (A'Hearn et al. 1995, Kidger 2004, Schleicher 2006, Agarwal et al. 2007) found a wide range of exponent values for the afρ decrease of 67P post-perihelion, i.e. from γ = -0.4 to γ = -9.5. Our γ exponent values for a likely continuous dust activity decrease are -4.1 ro -4.2 (for the linear – β = 0.04 - and HG-type phase functions – G = 0.15), which is considerably steeper than the one of Snodgrass et al. (2013) with γ ~ -3.4. It is noted that phase reddening effects of the 67P dust are not obvious from our observations (see Tab. 4) although our color data represent essentially only three epoch (10+12 September 2015, 29 January 2016 and 7/8 May 2016) with a phase angle difference of about 25 deg.

Additional indications for enhanced or reduced dust activity in 67P during the first quarter of 2016 do not exist in our data. Instead the afρ level in early May 2016 is in-line with a continuous drop in dust activity of 67P following the behavior as shown as of late September to December 2015. It is noted that also the fan structures seen in the dust coma of the comet do not indicate significant changes (see below), thus supporting a more continuous dust activity of 67P. Consequently, a linear (β=0.04) or HG-type (G=0.15) phase darkening describe best the post-perihelion brightness evolution of the comet for the time interval late September 2015 to early May 2016. The maximum afρ of about 1300 cm shortly after perihelion corresponds to about the same amount of dust production in kg/s, if one assumes the empirical calibration of A'Hearn et al. (1995) which for 67P is not yet verified. This estimation strongly depends on the adopted phase darkening of the comet and of course on various other properties of the coma dust and may at best provide only an order of magnitude reference.

*Radial coma flux profiles:* The overall status of the dust coma is assessed by measuring the radial flux profile of the comet in concentric circular apertures, centered on the brightness peak in the coma. WWFI r filter images of 67P are chosen for this analysis, since they are usually exposed longest and show best the dust distribution in the coma. Moreover, they are less or only very weakly contaminated by coma gas and are by far most abundant compared to WWFI i images in our dataset. Figure 6 shows examples of the radial coma profiles measured throughout the observing campaign. It is noted that for all observing epochs the profile slope becomes flatter with increasing distance from the nucleus. Due to radiation pressure more sensitive, lighter and smaller grains are pushed stronger into anti-solar direction such that they are missing in the size distribution at larger distances from the nucleus, in particular in the direction to the Sun. Hence, the radial profile gets flatter for larger distance. For a quantitative assessment of the profile, the exponent of the radial flux decrease is fitted to the measured radial profile for different distance ranges in the coma, i.e. log flux versus log aperture radius for 5000 to 10000 km, for 10000 to 15000km, for 15000 to 20000 km and for 20000 to 25000 km projected distance in the cometary coma. The results are tabulated in Tab. 3 and plotted in Fig. 7.

From the end of September to at least the end of December 2015 the dust coma of 67P appeared to be globally in 'quasi-steady-state', since the radial exponents of the flux profiles were close to unity. As of late January 2016 and more pronounced during the subsequent months, the radial exponent of the measured inner most coma region (5000 to 10000 km) – as well as the outer ones - increased above 1. The dust trail in the aperture radius, although at higher relative level with respect to the coma light than closer to perihelion, has a shallower brightness gradient than measured in the overage coma profile and is thus not responsible for the steeper gradients in the average coma profiles. This indicates that the 'quasi-steady-state' period for the coma dust had ended. With increasing solar distance the dust leaving the gas-dust coupling zone around the nucleus has slower initial expansion velocity into the coma. Then, its traveling time for a passage through the inner most measurement aperture amounts to several days to a few weeks, meaning dust is piling up in the inner part of the coma, while the outer part contains dust released earlier and with higher expansion speed. As a consequence the radial slope in the coma profile increases above unity. A similar phenomenon, i.e. a very steep radial coma profile, was seen in 67P at solar distances around 3 AU, when the comet approached the Sun in 2008 (Tozzi et al. 2011). The authors explained the observations by a very steep drop of the dust expansion velocity with exponents for the solar distance dependence between 4.2 to 4.9.

**Figure 6** Normalized averaged radial coma profiles of 67P – changes with time. The figure plots for various dates during our observing campaign radial flux profiles from the coma center to about 38000 km project nucleus distance. The individual graphs show the average coma flux in the measurement aperture of increasing nucleus distance. For comparison the flux values are normalized to unity for aperture size 10000 km.

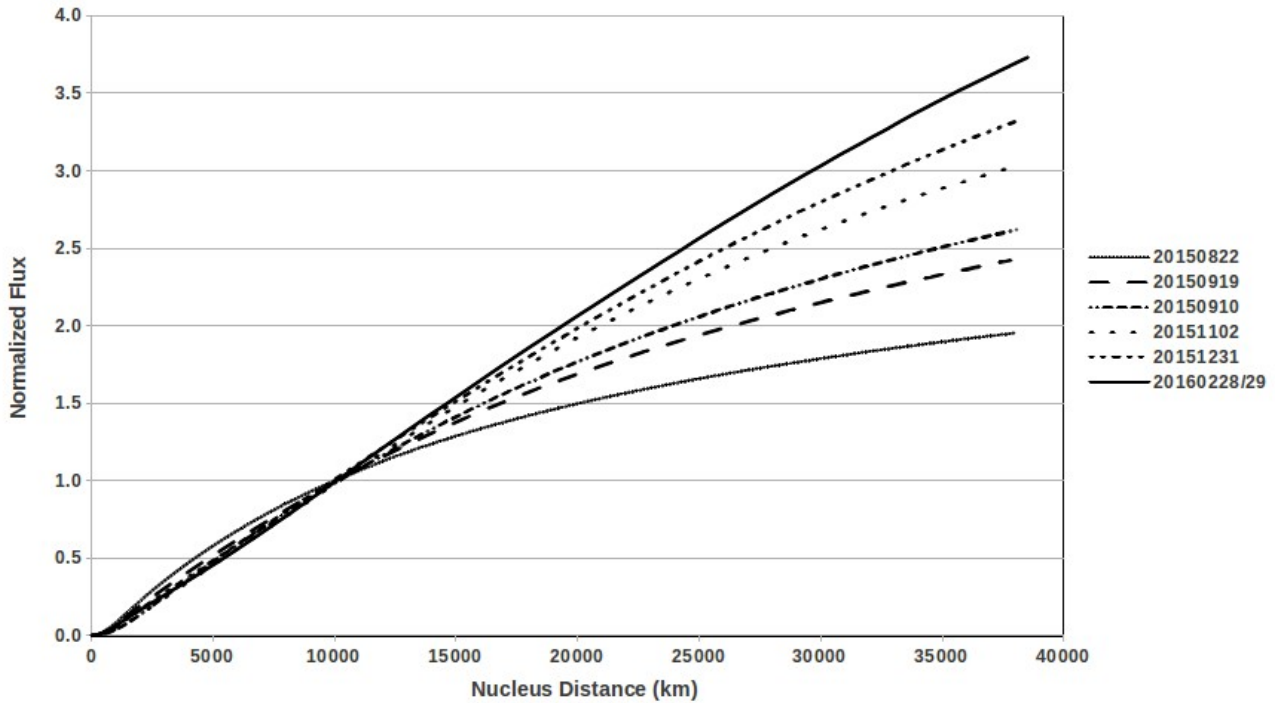

From 21 August to about mid September 2015 the exponent of the averaged radial dust coma profile is well below 1 for all distance ranges measured, reaching values of 0.8 to 0.45 at the beginning of the campaign. Ignoring albedo changes of the dust, this result implies that dust leaving the outer limit of the measurement aperture in the coma has a higher light scattering cross section than the one entering the inner edge. Generating additional light scattering cross section of the coma dust is possible if dust grains disintegrated throughout the coma of 67P. Hence, our measurements of the radial profiles in the coma of 67P suggest that dust fragmentation occurred within about one month after perihelion passage of the comet. Dust fragmentation by electrostatic forces (Boehnhardt&Fechtig 1987, Boehnhardt 1989) is unlikely since the charging conditions do not vary a lot with solar distance such that it should continue to affect the radial profile even at larger distances. Indeed, Fulle et al. (2015) argue for electrostatic fragmentation of dust when 67P was at larger distances from the Sun inbound. If this process still plays a role beyond the first month post-perihelion, it would be masked by the general evolution of the dust in the coma of 67P such that it is no longer discernible in ground-based observation alone. Boehnhardt&Fechtig (1987) have proposed an alternative fragmentation scenario based upon disintegration of larger grains into smaller building stones due to the evaporation of gluing material in larger aggregates. This process is likely temperature driven and thus depending on the solar distance of the comet. Another process that may cause grain disruption is aggregate bursting due to acceleration of the grain rotation by solar irradiation on irregular-shaped dust grains (windmill effect, Paddack 1969, Paddack&Rhee 1975). Despite that a quantitative assessment of the efficiency of this effect for the dust in 67P is difficult, it should result in lower efficiency with increasing solar distance of the comet.

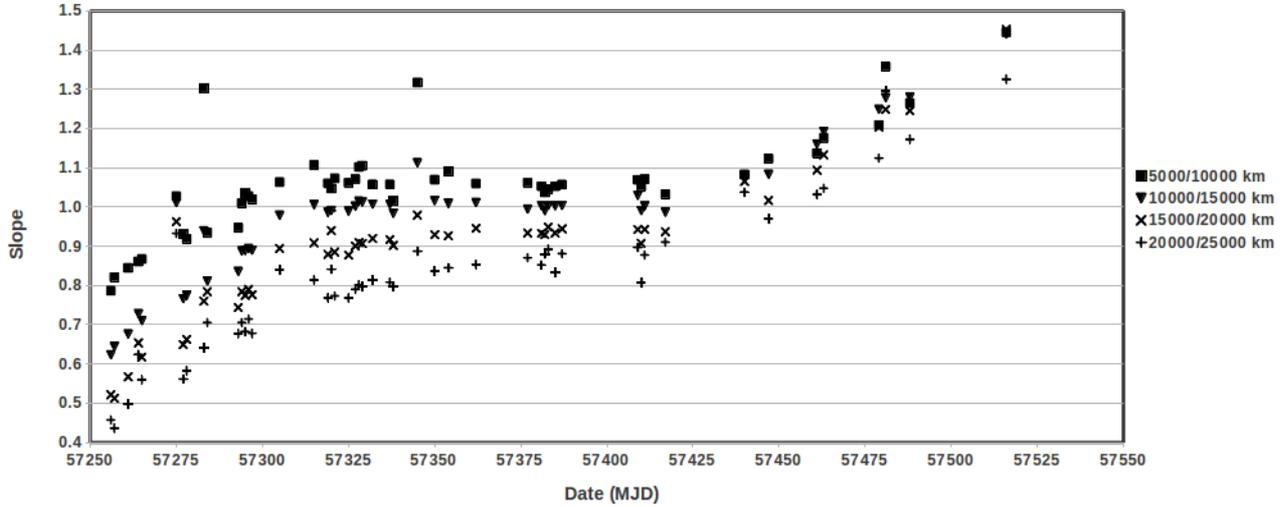

*F*igure 7 Radial slopes of the average coma profile of 67P versus time. The slope parameter is fitted to the profile in r filter images over different distance ranges in the coma, i.e. between 5000 to 10000 km, 10000 to 15000 km, 15000 to 20000 km and 20000 to 25000 km. Outlier values (on 19 September 2015, 19 November 2015, 3 April 2016) are due to bad seeing (3 arcsec and more) conditions.

*Cometary tail:* Over the whole observing period the tail of 67P appears as a narrow fuzzy extension in the north-western quadrant of the images, as of late October with a sharply peaked embedded axis of maximum brightness. It extends beyond the edge of the field of view of our exposures several 100000 km away from the nucleus (Fig. 8). The tail is more diffuse towards the southern edge and at the comet head it connects to the coma isophotes.

In the Wendelstein images of 67P narrow linear and kinked structures, indicative for ionic features of the plasma tail, are not identified. We thus believe that the tail appearance in the images is produced by dust reflected sunlight. The position angles (PA) of the main tail axis (PTA), represented by the line of maximum brightness is measured in the images close to the nucleus position and are tabulated in Tab. 6 together with the PAs of the extended radius vector of the comet (PSANG) and that of the negative direction of the comet's heliocentric velocity vector (PSAMV), both projected onto the sky of the observer. PSANG is usually indicative for the tail region containing 'young' dust, released days to a few weeks before the observing epoch, as well as for the approximate direction of the ion tail. PSAMV is more representative for the location of older dust grains in the tail, released months in the past. Tab. 6 provides also information on position angles for so called dust synchrones in the tail region. Synchrones are location of dust grains of different solar radiation pressure sensitivity, but released at the same time from the nucleus. The synchrone PAs of 67P were estimated from Finson-Probstein calculations of its dust tail (Finson&Probstein 1968; Beißer&Boehnhardt 1987). The table lists the synchrone PAs for dust, released 10 days before observing epoch (labeled DT=10), thus representing the young dust in the tail, for dust released at perihelion (labeled T=0) representing the period close to highest activity of the nucleus, and for dust released 400 days before perihelion (labeled T=-400), representing old dust emitted from the nucleus when the comet was beyond 3 AU from Sun inbound. Since the changes of PAs are relatively slow with time, Tab. 6 shows results for a representative set of the available observing epochs, distributed over the whole observing interval of 67P from late August 2015 to early May 2016.

**Table 6** Position angle of the dust tail region of 67P, representing the observing interval of the comet from late August 2015 to early May 2016. The table columns list (from left to right): the observing date, the position angle of the extended radius vector of the comet (PSANG), the position angle of the negative direction of the comet's heliocentric velocity vector (PSAMV), the position angle PTA of the main tail axis measured in the images, the angles PTA-PSANG and PTA-PSAMV, the position angle of synchrones for dust released 10 days before observing epoch (PA DT=10), at perihelion passage (PA T=0) and 400 days before perihelion (PA T=-400), and the angular differences between PTA and the 3 synchrones, i.e. PTA-DT=10, PTA-T=0 and PTA-T=-400. PTA is measured to +/-2 deg, the other angles are accurate to +/-1 deg.

| Date (UT) | PSANG (deg) | PSAMV (deg) | Tail Axis PTA (deg) | PTA-PSANG (deg) | PTA-PSAMV (deg) | PA DT=10 (deg) | PA T=0 (deg) | PA T=-400 (deg) | PTA-DT=10 (deg) | PTA-T=0 (deg) | PTA-T=-400 (deg) |
|---|---|---|---|---|---|---|---|---|---|---|---|
| 20150822 | 278 | 269 | 281 | 3 | 12 | 276 | 277 | 270 | 5 | 4 | 11 |
| 20150823 | 279 | 269 | 283 | 4 | 14 | 277 | 277 | 270 | 6 | 6 | 13 |
| 20150827 | 281 | 272 | 284 | 3 | 12 | 279 | 278 | 272 | 5 | 6 | 12 |
| 20150830 | 282 | 273 | 283 | 1 | 10 | 280 | 279 | 273 | 3 | 4 | 10 |
| 20150831 | 282 | 273 | 283 | 1 | 10 | 280 | 279 | 273 | 3 | 4 | 10 |
| 20150910 | 286 | 277 | 290 | 4 | 13 | 284 | 283 | 277 | 6 | 7 | 13 |
| 20150912 | 287 | 278 | 290 | 3 | 12 | 285 | 284 | 278 | 5 | 6 | 12 |
| 20150919 | 289 | 281 | 294 | 5 | 13 | 286 | 285 | 281 | 8 | 9 | 13 |
| 20150930 | 292 | 285 | 295 | 3 | 10 | 290 | 288 | 285 | 5 | 7 | 10 |
| 20151010 | 294 | 288 | 293 | -1 | 5 | 293 | 290 | 287 | 0 | 3 | 6 |
| 20151020 | 295 | 290 | 297 | 2 | 7 | 294 | 292 | 290 | 3 | 5 | 7 |
| 20151101 | 296 | 293 | 300 | 4 | 7 | 295 | 294 | 293 | 5 | 6 | 7 |
| 20151112 | 296 | 296 | 303 | 7 | 7 | 296 | 295 | 295 | 7 | 8 | 8 |
| 20151124 | 296 | 298 | 304 | 8 | 6 | 295 | 296 | 297 | 9 | 8 | 7 |
| 20151206 | 295 | 300 | 305 | 10 | 5 | 295 | 297 | 298 | 10 | 8 | 7 |
| 20121221 | 294 | 303 | 305 | 11 | 2 | 294 | 297 | 300 | 11 | 8 | 5 |
| 20121229 | 293 | 304 | 302 | 9 | -2 | 293 | 297 | 300 | 9 | 5 | 2 |
| 20151231 | 292 | 304 | 303 | 11 | -1 | 293 | 297 | 300 | 10 | 6 | 3 |
| 20160122 | 288 | 305 | 304 | 16 | -1 | 289 | 297 | 301 | 15 | 7 | 3 |
| 20160130 | 286 | 305 | 305 | 19 | 0 | 287 | 297 | 300 | 18 | 8 | 5 |
| 20160222 | 273 | 303 | 300 | 27 | -3 | 277 | 295 | 300 | 23 | 5 | 0 |
| 20160228/29 | 261 | 301 | 300 | 39 | -1 | 270 | 295 | 299 | 30 | 5 | 1 |
| 20160316 | 170 | 300 | 300 | 130 | 0 | 233 | 294 | 299 | 67 | 6 | 1 |
| 20160401 | 131 | 298 | 296 | 165 | -2 | 136 | 294 | 295 | 160 | 2 | 1 |
| 20160410/11 | 124 | 297 | 295 | 171 | -2 | 126 | 294 | 295 | 169 | 1 | 0 |
| 20160507/08 | 116 | 296 | 297 | 181 | 1 | 116 | 295 | 295 | 181 | 2 | 2 |

The interpretation of the tail structure in the Wendelstein images is in good agreement with the time evolution of the measured position angles of the tail of 67P, as compared to PSANG and PSAMV (see Tab. 6). Since the orbit inclination of 67P is low (7 deg), the dust tail is mostly seen side-on with low out-of-plane viewing angles from Earth (below 5 deg). Only during the opposition time interval (January to May 2016) the tail region covers a wider range of PAs. From the comparison of the tail PAs measured (PTA) with the PSANG and PSAMV values as well as with the PAs of synchrones in Tab. 6 we conclude: Both, the young and the older dust tail overlap in the PA range spread by PSANG and PSAMV. From late August to about mid October 2015 the main axis of the dust tail close to the nucleus is defined by 'young' dust released within a few weeks before the observing epoch. At that time the tail is more diffuse which is assumed to be due to the higher out-of-orbital-plane velocity of the grains released closer to perihelion. A minor curvature to smaller PAs further away from the nucleus indicates the presence of 'older' (and larger) dust grains in the tail region, frequently associated with the dust trail of the comet. Around the period of the comet orbital plane crossing by the Earth (mid October to end November 2015; plane crossing happened on 13 November 2015) the young and the old dust fully overlap in the tail region. Thereafter, i.e. as of early December 2015 until the end of our observing period of the comet in early May 2016, the tail axis is associated with the 'older' dust, released several 100 or even more days before perihelion passage (13 August 2015). It is noted that the sweeping of the 'young' dust tail from the north-western to the south-eastern quadrant around comet opposition is not obvious from the Wendelstein exposures. In particular, the 'young' dust tail is – at best – only marginally detectable in anti-solar direction in April to early May 2016, despite an amplification of its scattering cross section due to another comet orbital plane passage by the Earth (10 May 2016). This may be due to the very much reduced activity of 67P beyond 2.5 AU from the Sun (see Fig. 5 and Tab. 3) as compared to the months following perihelion passage.

**Figure 8** 67P dust tail appearance and geometry during 3 nights over the observing interval. The panels (in isophote-like representation) show the average composite images of the WWFI r filter exposures of the respective nights. Stars and background objects appear as 'strings of pearls' in the images. Image orientation: North is up, East is to the left. Top panel: 10 September 2015, 20min total integration time, field of view 284000 x 156000 km; Middle panel: 11 November 2015, 52min total integration time, field of view 286000 x 228000 km; Bottom panel: 10-11 April 2016.,74min total integration time, 341000 x 252000 km field of view.

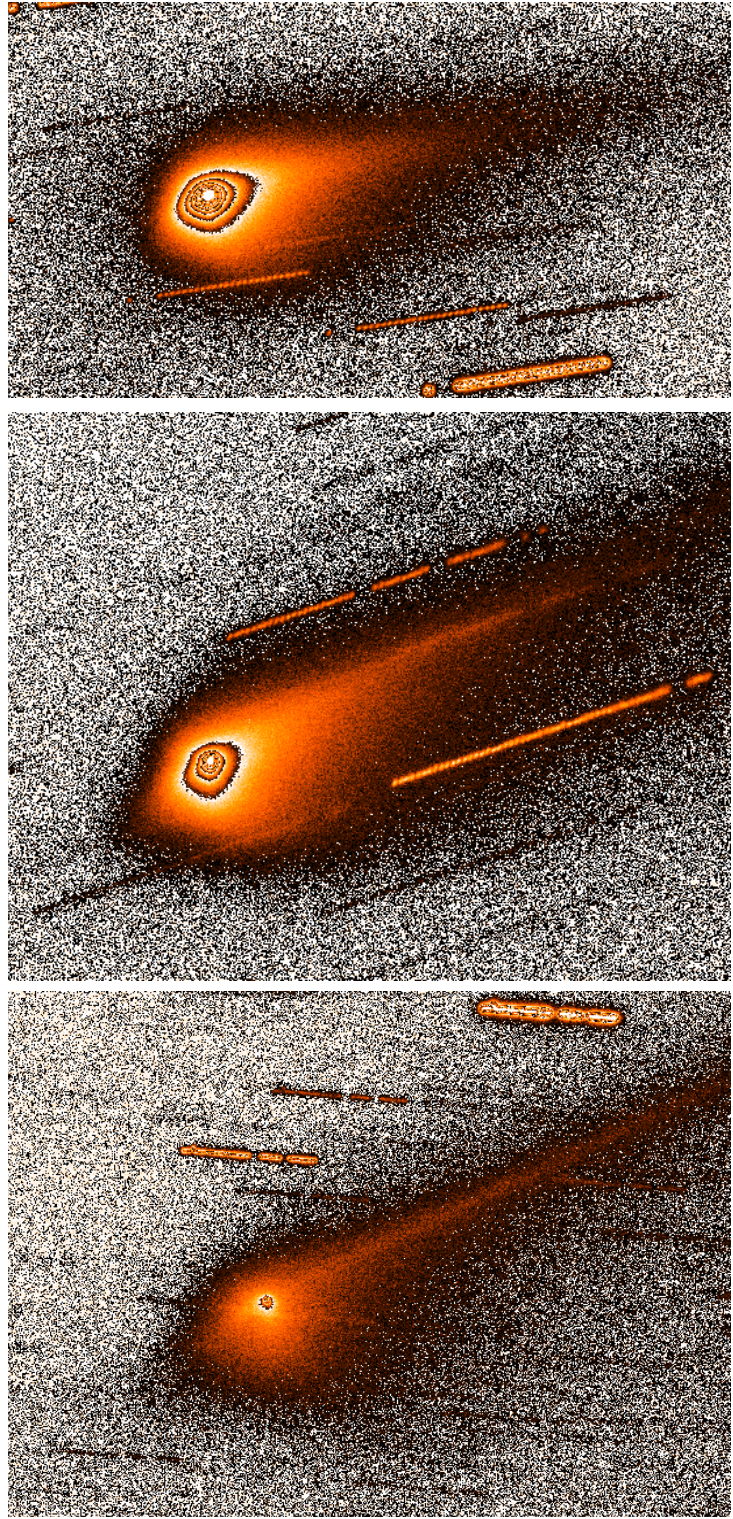

*Coma:* The dust coma of 67P is best depicted in the WWFI r filter images. The isophote pattern appears neither perfectly circular nor perfectly oval, which can be considered as a indicator for a rather uniform activity across the nucleus surface. Instead, it shows moderate 'bumps' indicative for localized coma features of enhanced dust reflectivity. The coma structure enhancement techniques applied to the images (i.e. radial renormalization and adaptive Laplace filtering, see section 3) reveal a coma pattern with 2 (pattern X1) or 3 (pattern X2) fan-like dust structures embedded in the bumpy isophote appearance of the comet - see Fig. 9 for the appearance of pattern X1 and X2 and Tab. 7 for characteristic information of the fan structures in the patterns.

Pattern X1 has a fan (labeled B in Tab. 7 and Fig. 9) of about 15 to 30 deg opening angle, pointing approximately in east-south-eastern direction, and another fan (labeled C in Tab. 7) of 10 to 20 deg opening angle, pointing approximately to the South. Both fans B and C vary slowly with time in their near-nucleus position angles and in their approximate brightness ratio relative to the mean coma brightness, estimated at about 5000 km distance from the nucleus. The enhanced images (for examples see Fig. 9) allow the measurement of the near-nucleus position angles of the fan-like coma structures and they also show their further extension in the coma out to several 10000 km projected nucleus distance. Straight and positively curving (increasing PA counted East over North) fans are noticed.

Pattern X2 contains the two fan-like structures B and C of pattern X1 and in addition a short, about 10-20 deg wide fan A, pointing in approximately east-north-eastern direction (see Tab. 7 and Fig. 9). Pattern X1 was observed from 22 August (earliest observing date at Mt. Wendelstein for 67P) until late October 2015. Pattern X2 followed after X1 and lasted at least until mid April 2016.

In the following, we describe the individual fan-like dust structures in the coma and assess the possible origin for the dust emission. For the latter, we apply dust coma modeling after Beisser&Drechsel (1992) which is based on a Monte-Carlo Finson-Probstein algorithm for dust, emitted from different locations on a rotating spherical nucleus. Since the nucleus of 67P has a bi-lobate shape (Sierks et al. 2015), our modeling results cannot adequately represent the dust emission geometry on the nuclear surface, but they are taken as an approximation for the dust flow geometry at larger distances (>1000km, the typical size of the seeing disk in ground-based observations) from the nucleus beyond the dust-gas coupling and acceleration zone. For the rotation motion of the nucleus the axis orientation and rotation period given by Sierks et al. (2015) are applied. Since – by the time of writing - the Rosetta images of the 67P dust and gas activity close to the nucleus for the time period August 2015 to May 2016 are not yet publicly available, a correlation assessment of structures seen in ground-based and Rosetta images was not performed and remains to be done in the future. It is noted that the appearance of the individual fan-like structures described below did not change noticeably during the observing intervals of the individual observing nights. This indicates that the nucleus rotation did not modulate the fan-like coma structures on time scales of hours, i.e. during a single nucleus rotation.

*Fan A:* Fan A evolves only as of early November 2015. It occurred first at about 60 deg position angle as a narrow, straight and several 1000km long feature. With time it became more extended and it converged towards eastern direction, keeping always a straight appearance. It reached maximum intensity ratios of up to 1.15 compared to the average coma profile at 5000 km distance. The fan modeling suggests a near-equatorial location for the dust emitting region of fan A.

*Fan B:* This fan structure resided in the south-eastern quadrant of the coma during the whole observing period. The opening angle shrank after about orbital plane passage by the Earth around mid November 2015. The fan curvature was positive (counter-clockwise) until November 2015, thereafter it got rather straight until the end of the observing interval. It is noted that during the first two months of the observing period the tail ward curvature of the dust in the fan was rather sharp at about 25000 km in the sun ward coma hemisphere, suggesting the presence of dust grains of expansion velocity up to 50-100 m/s for a solar radiation pressure sensitivity of up to 0.5. Up to the time of orbital plane passage by the Earth (13 November 2015) a short northward extension of the fan evolved a few thousand km projected distance from the coma center, indicating dust grains expanding into low obliquity directions towards the North of the sub-solar latitude. The Monte-Carlo modeling suggests that fan B is related to dust emission from mid-southern latitudes (40-50 deg South). The short northward extension of the fan could result from activity from low southern latitudes, possibly even from the source regions of fan A. This implies that the fan activity started at relatively high solar illumination angles of the source region.

*Fan C:* This fan-like structure is visible in southern direction over almost the whole observing period when 67P images were taken, i.e from late August 2015 until early April 2016. It is not identified in the May 2016 images; however, these exposures have only low signal-to-noise ratios such that it may have escaped detection. The opening angle is about 15-20 deg and it reached highest brightness ratios around October to November 2015. Fan structure C showed positive curvature. However, it displayed a straight figure in the near-nucleus region during the first 2 ½ months of our observations of 67P (up to late October 2015). It turned into tail direction at about 40000 km projected nucleus distance. The fan modeling suggests that coma structure C is produced by activity in high southern latitudes (most likely 60-80 deg South) on the nucleus. The opening angle requires either a certain latitude and/or longitudinal extension or a significantly long daily activity cycle of the related surface region. The curvature of the fan C suggests expansion velocities in the range of 50-100 m/s for dust grains of a solar pressure sensitivity of up to 0.5.

*Dust arc on 23 August 2015:* Comparison of Laplace filtered images of 67P, obtained on 22 and 23 August 2015, shows a temporary structure in the northeastern coma quadrant (Fig. 10). The structure consists of a narrow (< 1000 km wide), linear jet-like feature, extending at position angle of about 20 deg up to projected nuclear distances of about 13000 km. At that location a ~2500 km wide arc-like structure connects to the jet, pointing at position angle of about 45 deg and extending for another 11000 km. Both jet and arc are detected on 23 August 2015, but they are not seen in 67P images taken on 22 and 26 August 2015, the two observing epochs in our dataset closest in time to 23 August 2015. The temporary pattern on 23 August 2015 results also in more extended isophotes in the respective position angle range (Fig. 10, top panels) which could be understood as being due to an enhanced amount of dust in this part of the coma compared to the day before. However, since only WWFI r filter images are taken on 23 August 2015, a gaseous origin of the temporary jet- and arc-like structures cannot be excluded, but appears to be unlikely given the rather confined geometry of both features. As can be seen in Fig. 1 and Tab. 1, a small brightness increase of about 0.1mag is noted on 23 August 2015 compared to the coma brightness of the day before. However, since this amplitude is at the level of the estimated intrinsic average accuracy of our coma photometry and since a very similar brightness level of the comet is measured on 26 August 2015, we do not have a clear evidence for a relationship between the measured photometric brightness increase and the appearance of the arc and jet structure. The ratio image between 23 and 22 August 2015 shows a local brightness increase in the northeastern coma quadrant that reaches a level of 1.7 and 1.3 in the regions of the shell and jet structure, respectively.

The connecting geometry of arc and jet suggest a physical relationship for their origins. Dust activity modeling suggests that the arc material expands from low positive latitudes between about +5 to +10 deg. The similar projected distance of the arc from the nucleus and its uniform width can be understood by dust production of the arc material peaking at about the same time over a short time interval and over a latitude range of about 5 deg on the nucleus. Assuming that the dust emis-sion event may have started 24 hours earlier on 22 August 2015 (when the jet and arc structure are not seen in our images), the arc material may have expanded with a velocity of about 150 m/s and a velocity spread in the order of 30 m/s. The event must have been very short in time (an ejection activity peak of the order of half hour or less) in order to allow for the confined arc appearance 24hours later. The jet-like structure seen in the images of 23 August 2015 suggests that the activity at the northern edge of the ejection region continued for at least 24h. Up to now it was not possible to associate unusual instrument measurements on-board Rosetta with the origin of the arc and jet-like features, since the mission data are not yet publicly available (and the orbiter instruments have not fully taken note of the event seen in the Wendelstein data despite that it was reported to the project within a few days after the event). However, for a arc-jet scenario as outlined above, detection by Rosetta in-situ and remote sensing instruments requires the orbiter to fly – in a few 100 km distance - across the proper surface area at the right moment very close in time when the event happened or to image the respective region of dust ejection at the right moment (order of a few minutes).

**Table 7** Characteristics of the 67P dust coma fans. The table columns list (from left to right): the date of the observation (UT), the position angle of the extended radius vector of the comet (PSANG in deg), the position angle of the negative direction of the comet's heliocentric velocity vector (PSAMV), the sub-solar latitude on the nucleus of 67P (south = negative; deg), the near-nucleus position angle of a fan produced by an active region on the southern rotation pole of the nucleus (deg), the fan pattern label (X1 or X2 as described in the text), the near-nucleus position angles of fan structures A, B, C (measured North over East; deg), the near-nucleus opening angle of the fan-like structures A, B, C, the overall curvatures of the fan-like structures (s = straight, p = counter-clockwise, n = clockwise) and the intensity ratio of the fan maximum brightness at about 5000km projected distance to the nucleus compared to the average coma brightness level at that distance for fan-like structures A, B, C. NM indicates non-measurable property of the fan-like structures, ? identifies a tentative, but uncertain estimation.

| Date (UT) | PSANG (deg) | PSAMV (deg) | Subsolar Latitude (deg) | Angle Rotation axis (deg) | Pattern Type | Position Angle A (deg) | Position Angle B (deg) | Position Angle C (deg) | Opening Angle A (deg) | Opening Angle B (deg) | Opening Angle C (deg) | Curvature A | Curvature B | Curvature C | Intensity ratio A At 5000 km | Intensity ratio B At 5000 km | Intensity ratio C At 5000 km |
|---|---|---|---|---|---|---|---|---|---|---|---|---|---|---|---|---|---|
| 20150822 | 278 | 269 | -51 | 165 | X1 | NM | 101 | 188 | NM | 30 | 15 | NM | p | sp | NM | 1.3 | 1.2 |
| 20150823 | 279 | 269 | -51 | 165 | X1 | NM | 103 | 189 | NM | 30 | 15 | NM | p | sp | NM | 1.3 | 1.2 |
| 20150827 | 281 | 272 | -52 | 165 | X1 | NM | 104 | 189 | <10 | 30 | 15 | NM | p | sp | NM | 1.3 | 1.25 |
| 20150830 | 282 | 273 | -52 | 164 | X1 | NM | 103 | 190 | NM | 30 | 15 | NM | p | sp | NM | 1.3 | 1.25 |
| 20150831 | 282 | 273 | -52 | 164 | X1 | NM | 103 | 190 | NM | 30 | 15 | NM | p | sp | NM | 1.3 | 1.25 |
| 20150910 | 286 | 277 | -52 | 160 | X1 | NM | 114 | 166 | <10 | 30 | 15 | NM | p | sp | NM | 1.3 | 1.25 |
| 20150912 | 287 | 278 | -52 | 159 | X1 | NM | 114 | 168 | NM | 30 | 15 | NM | p | sp | NM | 1.3 | 1.25 |
| 20150919 | 289 | 281 | -51 | 159 | X1 | NM | 116 | 172 | NM | 30 | 15 | NM | p | sp | NM | 1.3 | 1.25 |
| 20150930 | 292 | 285 | -48 | 158 | X1 | NM | 118 | 175 | NM | 30 | 20 | NM | p | sp | NM | 1.35 | 1.3 |
| 20151010 | 294 | 288 | -44 | 156 | X1 | NM | 118 | 175 | NM | 25 | 20 | NM | p | sp | NM | 1.4 | 1.3 |
| 20151020 | 295 | 290 | -41 | 156 | X1 | NM | 123 | 173 | <15 | 25 | 20 | NM | p | sp | NM | 1.4 | 1.35 |
| 20151101 | 296 | 293 | -36 | 155 | X2 | 60 | 128 | 175 | 30 | 20 | 20 | s? | p | sp | NM | 1.4 | 1.4 |
| 20151112 | 296 | 296 | -32 | 158 | X2 | 63 | 125 | 180 | 30 | 20 | 20 | s | p | sp | NM | 1.35 | 1.4 |
| 20151124 | 296 | 298 | -28 | 158 | X2 | 65 | 130 | 183 | 30 | 20 | 20 | s | p | sp | NM | 1.30 | 1.35 |
| 20151206 | 295 | 300 | -24 | 159 | X2 | 70 | 128 | 190 | 30 | 15 | 20 | s | sp | sp | NM | 1.2 | 1.3 |
| 20121221 | 294 | 303 | -20 | 160 | X2 | 74 | 126 | 189 | 15 | 15 | 15 | s | sp | s | NM | 1.15 | 1.15 |
| 20121229 | 293 | 304 | -17 | 161 | X2 | 77 | 120 | 196 | 15 | 15 | 15 | s | sp | s | 1.05 | 1.25 | 1.15 |
| 20151231 | 292 | 304 | -17 | 161 | X2 | 74 | 123 | 199 | 15 | 15 | 10 | s | sp | s | 1.1 | 1.15 | 1.1 |
| 20160122 | 288 | 305 | -11 | 161 | X2 | 74 | 120 | 200 | 15 | 15 | 10 | s | s | s | 1.1 | 1.15 | 1.05 |
| 20160130 | 286 | 305 | -9 | 161 | X2 | 77 | 121 | 202 | 15 | 15 | 10 | s | s | s | 1.1 | 1.15 | 1.05 |
| 20160222 | 273 | 303 | -5 | 160 | X2? | NM | 118 | NM | NM | 15 | NM | NM | NM | NM | NM | 1.1 | NM |
| 20160228/29 | 261 | 301 | -3 | 159 | X2 | 77 | 117 | 220? | 15 | 15 | 15 | s | sp | s | 1.15 | 1.15 | 1.05 |
| 20160316 | 170 | 300 | -1 | 157 | X2 | 78 | 117 | 210 | 15 | 15 | 15 | s | sp | s | 1.15 | 1.2 | 1.1 |
| 20160401 | 131 | 298 | 2 | 156 | X2 | 80 | 115 | 205 | 10 | 20 | 20 | s | s | sp | 1.1 | 1.2 | 1.15 |
| 20160410/11 | 124 | 297 | 3 | 156 | X2 | 75 | 108 | 202 | 10 | 20 | 20 | s | s | s | 1.1 | 1.2 | 1.15 |
| 20160507/08 | 116 | 296 | 7 | 156 | X2 | 75 | 105 | NM | NM | NM | NM | s | s | NM | NM | NM | NM |

**Figure 9** Patterns X1 and X2 of fan-like structures in the coma of 67P. The top row shows the radially renormalized ratio (left) image and the Laplace filtered image (right) of 67P on 30 September 2015 as a example of the X1 pattern in the coma. Image scale is 130000 x 130000 km at the comet. The bottom image pair provides the respective example image for the X2 pattern in the coma as observed on 16 March 2016. Image scale is 58000 x 58000 km at the comet. For all images: North is up and East to the left. The coma fans are labeled by letters A, B , C. The dust tail is indicated by letter T.

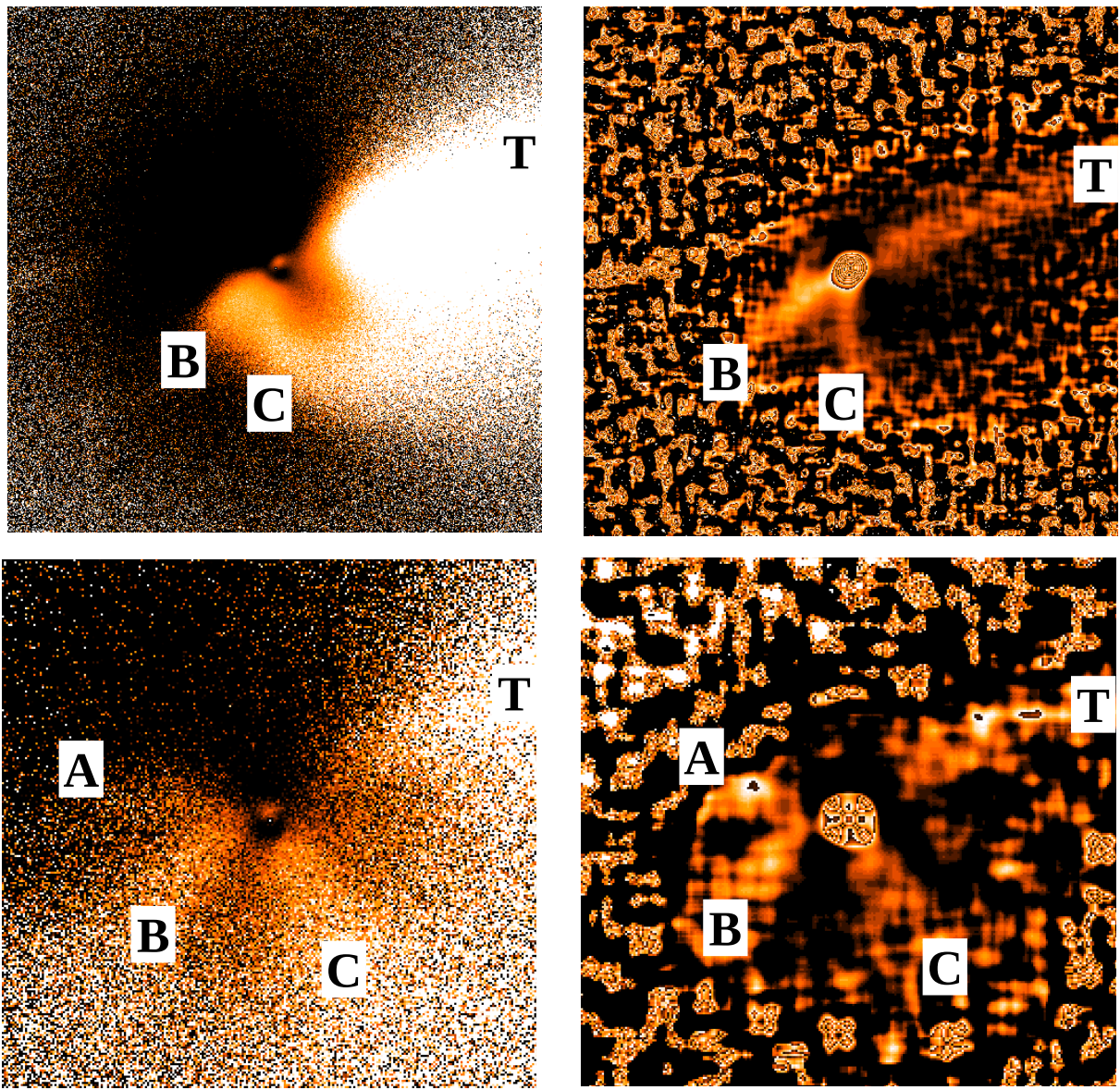

**Figure 10** Dust arc in the coma of 67P. The images show the isophote pattern of the comet on 22 August 2015 (top left panel), the isophote pattern of the coma on 23 August 2015 (top right panel), and the respective adaptive Laplace filtered versions of the comet image on both observing epochs (bottom left panel for 22 August 2015, bottom right for 23 August 2015). The isophote pattern on 22 and 23 August 2015 differ in the northeastern coma quadrant. The Laplace filtered image of 23 August 2015 displays a straight short jet-like structure at near-nucleus position angle 20 deg and an arc-let structure at the end of the jet, extending into southwestern direction. Both features are not present in the coma the day before. Image orientation is: North up, East to the left, image scale is 63000 x 61000 km at the distance of the comet.

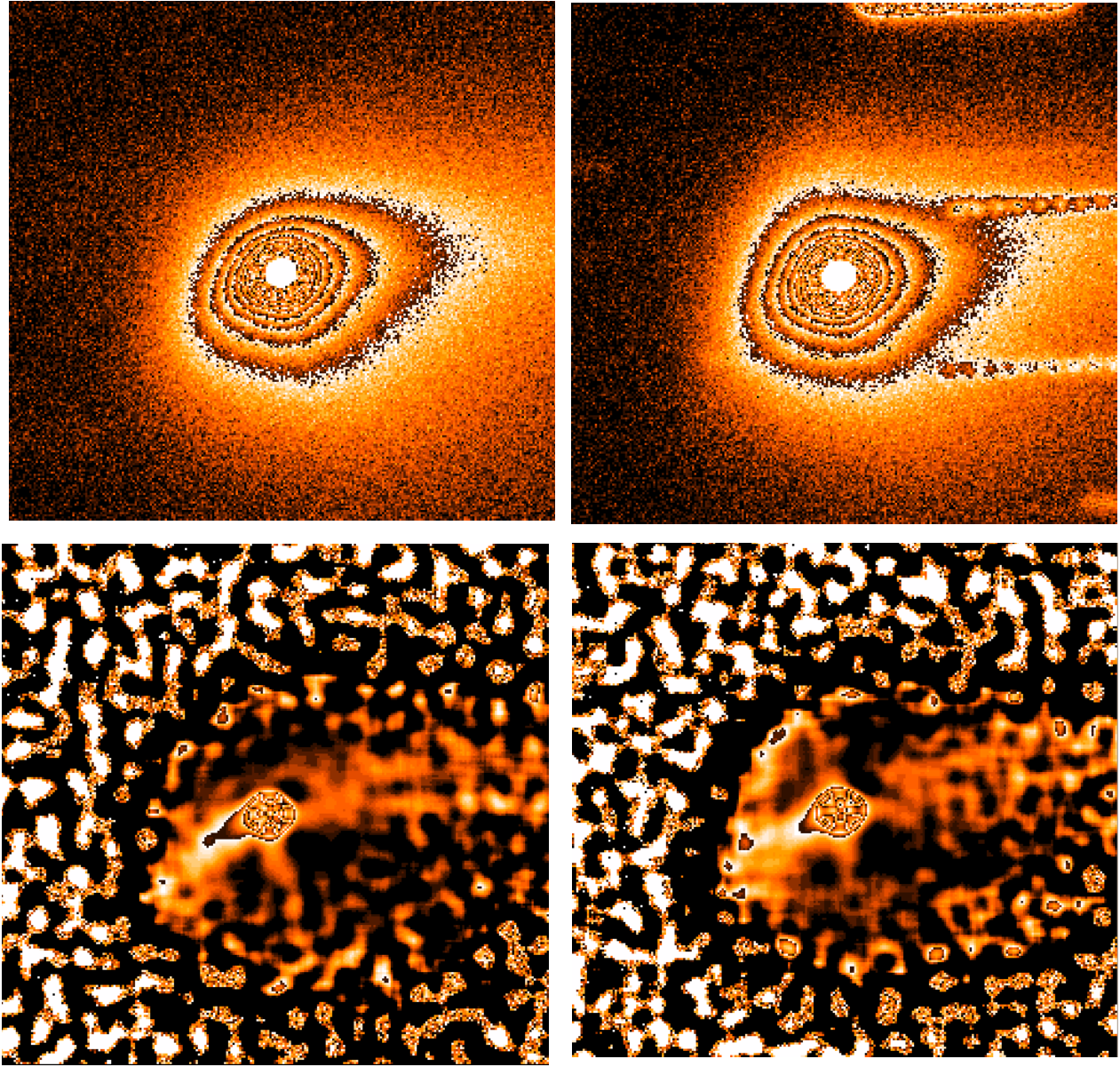

## 5. Conclusions and Discussion

Comet 67P was observed from the Mt. Wendelstein observatory by broadband imaging from 22 August 2015 to 9 May 2016. The images through the WWFI r filter show the dust coma and tail of the comet between 9 to 270 days after perihelion passage on 13 August 2015. Indications for the gaseous coma are noted for 2 observing epochs (10 and 12 September 2015) from color difference in the g-r and g-i versus the r-i wavelength range. The dust color corresponds to a moderately red spectral slope of about 8.5 %/100nm for the WWFI r and i filter wavelength range. The dust activity of 67P, represented by the af$\rho$ parameter obtained from the measured coma brightness in the r filter, reached its maximum during the first weeks of our observing campaign and it dropped steeply and continuously as of the 2nd half of September 2015 until the end of our campaign. A plateau in the af$\rho$ curve versus solar distance from the end of January to April 2016 is interpreted as being due to the opposition surge of the light scattering phase function. A linear (with slope 0.04) or asteroid-like HG-type phase function (with G ~0.15) describes a uniform activity decrease of the comet with heliocentric distance exponent of 4.1 – 4.2. However, from our observations a longer lasting activity bump or even a drop in the activity around opposition cannot be fully excluded depending on the dust phase function assumed.

The averaged brightness coma profiles displayed close to unity radial gradients from the end September until late December 2015 indicating a 'quasi-steady-state' dust coma. Before that time period the radial slope of the coma brightness was below unity which is interpreted as being due to the production of additional light scattering cross section with nuclear distance, possibly from dust aggregate fragmentation throughout the coma. In 2016 the averaged radial brightness profile of the coma got steeper than unity which might be a consequence of a 'non-steady-state' coma due to very much reduced dust expansion speed and thus an accumulation of dust activity in the inner coma. In this context it is noted that for a dust expansion speed of 50 to 100 m/s, which may be considered representative for the time close to perihelion, the traveling time of dust grains through a 10000 km radius measurement aperture is about 1-2 days. For slow changes in the dust activity versus time as it is seen in 67P, the coma is considered to be in 'quasi-steady-state'. For expansion speeds reduced by a factor 10 – which may be typical for the comet at larger distances from the Sun – the dust traveling time through the 10000 km measurement aperture is in the order of 10-20 days, a situation when the coma is no longer in 'quasi-steady-state'. This also implies that the af$\rho$ values are to be taken as a kind of average value over a longer period of dust activity when changes of dust production and other dust parameters might occur and might impact the af$\rho$ result.

The dust tail appearance changed from a normal tail geometry before opposition to an anti-tail one after opposition in early March 2016. Close to perihelion the tail of 67P reflects mostly the distribution of younger dust overlapping the distribution of older dust. Around the time of orbital plane passage by the Earth in November 2015 the older dust with grains produced more than a year before perihelion passage dominated the tail appearance until the end of the observing campaign in May 2016.

Two patterns of coma structures are observed in 67P, i.e one up to the end of October 2015 showing 2 fan-like dust features in the southern and eastern coma quadrants and one as of thereafter showing - in addition to the two others - a fan structure in the east-north-eastern part of the coma. The three fans are explained as being due to dust activity in high (southern fan), mid (eastward fan) and near equatorial (east-north-eastern structure) regions of the nucleus. The latitude ranges found for dust activity in the fan-shape coma are consistent with results published by Vincent et al. (2013) from ground-based imaging before the Rosetta encounter with the comet. Both our modeling and the one performed by Vincent et al. apply a spherical nucleus shape and are thus not necessarily representative for a final solution using the bi-lobate shape and actual surface geometry of the nucleus of 67P. The association of near-nucleus cometary activity from Rosetta with the large-scale picture of the coma as seen for instance in Earth-based observations still has to be evaluated. The fan-shape coma as observed from Earth has to be traced back to the near-nucleus dust emission pattern as observed by Rosetta. As such our analysis results represent at least approximately the dust expansion beyond a few 1000 km nuclear distance in the coma.

The global picture of 67P after perihelion passage 2015 resembles in many respect the appearance during earlier apparitions of the comet, i.e. the dust activity (af$\rho$ versus time) and the fan-shape coma (see for instance Snodgrass et al. 2013, 2016 and references therein). This suggests that the activity pattern of 67P may not have changed much over the time period of a revolution around the Sun. This may apply as well for the rotation motion of the nucleus and the location of regions of enhanced dust emission from the nucleus.

An isolated dust ejection happened on the nucleus on 22-23 August 2015 resulting in a arc and jet-like feature in the northwestern coma quadrant. The event may have been of very short duration, expanding dust at high speed (order 100 m/s) from a region at low, but extended positive latitudes (5-10deg). An association of near-nucleus localized dust activity as measured by Rosetta and that seen in ground-based images was not yet attempted.

**Acknowledgement**

The authors of this paper want to thank Dr. J. Koppenhoefer for his support in the provision of the Pan-STARRS1 catalog data and Dr. K. Beisser for making available the Monte-Carlo simulation software for the dust coma and tail of comets. We also wish to acknowledge the important contributions for this work resulting from the Pan-STARRS1 Survey catalog data. The Pan-STARRS1 Surveys (PS1) have been made possible through contributions of the Institute for Astronomy, the University of Hawaii, the Pan-STARRS Project Office, the Max-Planck Society and its participating institutes, the Max Planck Institute for Astronomy, Heidelberg and the Max Planck Institute for Extraterrestrial Physics, Garching, The Johns Hopkins University, Durham University, the University of Edinburgh, Queen's University Belfast, the Harvard-Smithsonian Center for Astrophysics, the Las Cumbres Observatory Global Telescope Network Incorporated, the National Central University of Taiwan, the Space Telescope Science Institute, the National Aeronautics and Space Administration under Grant No. NNX08AR22G issued through the Planetary Science Division of the NASA Science Mission Directorate, the National Science Foundation under Grant No. AST-1238877, the University of Maryland, and Eotvos Lorand University (ELTE). This research was supported by the cluster of excellence 'Origin and structure of the universe' of the Deutsche Forschungsgemeinschaft DFG.

# References

Agarwal J., Müller M. Grün E., 2007, Space Sci. Rev., 128, 79
Agarwal J, M. Müller M., Reach W.T. et al., 2010, Icarus, 207, 992
A'Hearn M.F., Millis R.L., Thompson D.T., 1984, Icarus, 55, 250
A'Hearn M.F., Hoban S., Birch P.V. et al., 1986, Nature, 324, 649
A'Hearn M.F.,Millis T.L., Schleicher D.G., Osip D.J., Birch P.V., 1995, Icarus, 118, 223
Beisser K., Boehnhardt H., 1987, ApSS, 139, 5
Beisser K., Drechsel H., 1992, ApSS, 191, 1
Boehnhardt H., Fechtig H., 1987, A&A, 187, 824
Boehnhardt H., 1989, EMP, 46, 221
Boehnhardt H., Birkle K., 1994, A&A Suppl., 107, 101
Bowell E., Hapke B., Domingue D. et al., 1989, in: Asteroids II (eds. Binzel R.P., Gehrels T., Matthews M.S.), Univ. Arizona Press, 524
Cremonese G., Simioni E., Raggazoni R. et al. 2015, A&A 588, A59
Doressoundiram A., Boehnhardt H., Tegler S.C., Trujillo C., 2008, in: The Solar System Beyond Neptune (eds. Barucci M.A., Boehnhardt H., Cruikshank D.P., Morbidelli A.), Univ. Arizona Press, 91
Ferrin I., 2005, Icarus, 178, 493
Finson M.J., Probstein R.F., 1968, ApJ, 154, 327
Fornasier S., Hasselmann P.H., Barucci M.A. et al., 2015, A&A, 583, A30
Fukugita M., Ichikawa T., Gunn J.E. et al., 1996, AJ, 111, 1748
Fulle M., Della Corte V., Rotundi A. et al., 2015, ApJ, 802, L12
Gössl C., Riffeser A., 2002, astro-ph/0110704
Gössl C., Bender R., Fabricius M. et al., 2012, SPIE Proceedings, 8446, A84463P
Hadamcik E., Sen A.K., Levasseur-Regourd A.C. et al., 2010, A&A, 517, A86
Hopp U., Bender R. Grupp F. et al., 2014, SPIE Proceedings, 9145, A91452D
Ishiguro M., 2008, Icarus, 193, 96
Kelley M.S., Woodward C.E., Harker D.E. et al., 2006, ApJ, 651, 1256
Kelley M.S., Reach W.T., Lien D.J., 2008, Icarus, 193, 572
Kelley M.S., Wooden D.H., Tubiana C. et al., 2009, AJ, 137, 4633
Kidger M.R., 2004, A&A, 420, 389
Kosyra R., Gössl C., Hopp U. et al., 2014, Experimental Astronomy, 38, 213
Lamy O.L., Toth I., Fernandez Y.R., Weaver H.A., 2004, in: Comets II (eds. Festou M.C., Keller H.U., Weaver H.A.), Univ.Arizona Press, 223
Lamy P.L., Toth I., Weaver J.A. et al., 2006, A&A, 458,669
Lamy P.L., Toth I., Groussin O. et al., 2008, A&A, 489, 777
Lara L.M., de Leon J., Licandro J., Gutierrez P.J., 2005, EMP, 97, 165
Lara L.M., Lin Z.-Y., Rodrigo R., 2011, A&A, 525, A36
Lowry S., Fitzsimmons, A., Jorda, L., et al. 2006, in BAAS, 38, 492
Lowry S., Duddy S.R., Rozitis B. et al., 2012, A&A, 548, A12
Magnier E.A., Schlafly E., Finkbeiner D. et al., 2013, ApJS, 204, 20.
Müller, M., 1999, PhD Thesis, University of Heidelberg
Paddack S.J., 1969, JGR, 74, 4379
Paddack S.J., Rhee J.W., 1975, GeoRL, 2, 365
Schleicher D.G., Millis R.L., Birch P.V., 1998, Icarus, 132, 397
Schleicher D.G., 2006, Icarus, 181, 442
Schulz R., Stüwe J.A., Boehnhardt H., 2004, A&A, 422, L19
Sierks H., Barbieri C., Lamy P.L. et al., 2015, Science, 347, aaa1044
Snodgrass C., Tubiana C., Bramich D.M. et al., 2013, A&A, 557, A33

Snodgrass C., Jehin E., Manfroid J. et al., 2016, A&A, 558, A80
Tedesco E.F. 1989, in: Asteroids II (eds. Binzel R.P., Gehrels T., Matthews M.S.), Univ. Arizona Press, 1090
Tonry J.L., Stubbs C.W., Lykke K.R. et al. 2012, ApJ, 750, 99
Tozzi G.P., Patriarchi P., Boehnhardt H. et al., 2011, A&A, 531, A54
Tubiana C., Barrera L., Drahus M., Boehnhardt H., 2008, A&A, 490, 377
Tubiana C., Boehnhardt H., Agarwal J. et al. 2011, A&A, 527, A113
Vincent J.B., Lara L.M., Tozzi G.P., Lin Z.-Y., Sierks H., 2013, A&A, 549, A121
Weiler M., Rauer H., Helbert J., 2004, A&A, 414, 749